\documentclass[prd,nofootinbib,
amsfonts,amssymb,amsmath,twocolumn]{revtex4}
\usepackage{epsfig}
\usepackage{xcolor}
\usepackage{ulem}
\usepackage{appendix}
\begin{document}

\title{A frame-bundle formulation of quantum reference frames: 
\\
from superposition of perspectives to
superposition of geometries
}

\author{Daniel A.\ Turolla Vanzella}\email{vanzella@ifsc.usp.br}
\affiliation{Instituto de F\'\i sica de S\~ao Carlos,
Universidade de S\~ao Paulo, Caixa Postal 369, CEP 13560-970, 
S\~ao Carlos, SP,  Brazil}

\affiliation{Institute for Quantum Optics and Quantum Information (IQOQI), 
Austrian Academy of Sciences, Boltzmanngasse 3, A-1090 Vienna, Austria.\footnote{While on  sabbatical leave.}}

\author{Jeremy Butterfield}\email{jb56@cam.ac.uk}
\affiliation{Trinity College, University of Cambridge, Cambridge U.K.}

\begin{abstract}
Recent experimental advances suggest we may soon be able to probe the gravitational field of a mass in a coherent
superposition of position states---a system which is widely believed to lie outside the scope of classical and semiclassical gravity. The recent theoretical literature has applied the idea of {\it quantum reference frames} (QRFs), originally introduced for 
non-gravitational contexts, to such a scenario.

Here, we provide a  possible fully geometric formulation of  the core idea
of QRFs as it has been applied in the context of gravity, freeing its definition from unnecessary (though convenient) ingredients, such as coordinate systems. Our formulation is based on two main ideas. First, a QRF encodes uncertainty about 
 what is the observer’s (and, hence, the measuring apparatus's) perception of time and space  
 at each spacetime point (i.e., {\it event}). For this, 
 an observer at an event $p$ is modeled, as usual, 
 as a tetrad in the tangent space  $T_p$.
So a QRF  at an event $p$ is a complex function on the tetrads at $p$.  Second, we use the result that one can specify a metric on a given manifold by stipulating that a basis one assigns at each tangent space is to be a tetrad in the metric one wants to specify. Hence a spacetime, i.e.\ manifold plus metric, together with a
choice of ``point of view'' on it, is represented by a section of the bundle of bases, understood as taking the basis assigned to each point to be a tetrad. Thus a superposition of spacetimes gets represented as, roughly speaking, an assignment of complex amplitudes to sections of this bundle. 
A QRF, defined here as the collection of
complex amplitudes assigned to bases at events---i.e.,
a complex function defined on the bundle of bases of the 
manifold---can describe, in a local way (i.e., attributing the amplitudes to
bases at events instead of to whole sections), these superpositions.

 We believe that this formulation sheds some light on  some conceptual aspects and possible extensions of current ideas about QRFs.  For instance,
 thinking in geometric terms makes it clear that the idea of 
 QRFs applied to the
 gravitational scenarios treated in the literature
 (beyond linear approximation) 
  {\it lacks} predictive power due to arbitrariness which, we argue, can only be resolved by some further input from physics.

\end{abstract}

\maketitle

\section{Introduction}
\label{sec:intro}

Full comprehension of the interplay between gravity and the  principles which rule the quantum world
remains one of the
most elusive and challenging 
enterprises in  physics. In contrast to the situation
in the early decades of the twentieth century, when numerous puzzling
experimental results slowly, but steadily, 
paved the way towards an understanding of the
quantum 
aspects 
of matter and electromagnetism, the 
arid
landscape of observations  involving possible
quantum 
aspects of 
gravity has led
to the development of several different  
theoretical 
approaches to the subject but with almost no
guidance for us about what is the right direction.
To make things worse, our best current 
description of (classical) gravity, { viz.\ general relativity (GR)},
intermingles it with the 
most fundamental notions
of space and time, 
which apparently suggests the
need for a
radical change in the way we understand these fundamental
concepts if we are to succeed in this endeavor.

Recent advances in table-top experiments which are sensitive to
the gravitational field of tiny macroscopic masses (of order
$\lesssim 10^{-1}$~g~\cite{WHPA21}) 
have fostered the hope that in the
not-too-distant future we may begin probing 
the gravitational field of masses in quantum (coherent)
superposition of
position states. Although there is little doubt about what to expect
in these experiments in the linear regime of 
gravity---which is the regime we have some hope to probe---it is
important to explore the full implications of having
these coherent superpositions. Strictly speaking, the exact
analysis of
such a system
is already beyond our current understanding of
gravity and quantum physics.\footnote{Unless
it is eventually observed that the gravitational field of the coherent 
superposition 
is the same as the one generated by a {\it classical} mass 
distribution given by the average mass distribution of the
superposition, as described by the semiclassical
Einstein equation.}
Is it possible to have a 
(quantum) superposition
of geometries? What does it mean and what are 
its effects? These are  
some open
questions whose answers (even if tentative)  may take us
a step closer to a full understanding of the problem.

In an attempt to deal with situations such as the one described
above---quantum superpositions of well defined spacetimes---the idea of {\it quantum reference frames} (QRFs) has been
applied to these scenarios. 
It is important to clarify, at this point, that the term
 `QRF' has a long history in the literature (see, e.g., 
Refs.~\cite{AK84,BRS07,BRST09,PGB14}), not necessarily
 corresponding  exactly to the same idea---although some of them 
 may well be equivalent.
 Here, we take on some of its more recent uses as initiated by  Ref.~\cite{GCB19}:
 which was then
 applied to the
 gravitational scenario described above, such as 
 in Refs.~\cite{CGBB20,FB22,KHCB22,HKC23}.
Instead of focusing on 
what a superposition of  geometries {\it is} in general, 
QRFs 
 are used by these authors to address the much more pragmatic 
question of what are the {\it effects} of such a superposition on a 
(possibly quantum)
probe, in particular situations. 
In order to accomplish this,  these authors use
QRF {\it transformations}
and then invoke a 
 principle of ``quantum covariance.''
 
Using these ingredients, the authors of Ref.~\cite{HKC23}
calculate the evolution
of a ``test particle'' (i.e., a particle whose
own gravitational field is neglected in the analysis)
under the
influence of a mass in a macroscopic superposition of 
position states. In summary, this application of
QRFs formalizes, in a manifold with an undefined metric, the prediction that
the evolution of the test particle should be a 
superposition
of its ``histories,'' each  given by its evolution subject to
one well-defined position state of the source mass---at least 
as long as the test particle does not disturb the superposed mass state.

Our primary goal, here, is 
to recast this 
idea of QRFs and their
transformations
into a geometric language, avoiding
the use of auxiliary but unnecessary (and possibly 
confusing)
ingredients---such as
coordinate systems---in their constructions. 
 But in pursuing this goal, we are naturally led to
a framework where {\it arbitrary}  superpositions of {\it arbitrary} 
geometries
(on a given manifold) can be described in a local geometric (i.e., coordinate
independent) manner: a framework which may have some value, 
apart from our initial goal. But this extension beyond our initial goal is left largely unexplored in this paper.
% JBtoDV; Agree completely with your deletion and its reason,. so I have commented them out
%{\dv \sout{In particular, we will not consider how our framework could treat theories of gravity other than GR.}}
%{\dtoj [DVtoJB: This statement seems out of context because we are not really assuming GR but merely metric theories of gravity -- since we are nowhere making use of Einstein equations.]}
In this sense, we make no claim to give 
a complete, ``all-set'' { framework},
which leaves 
no open questions or loose ends. Rather, this paper is an attempt to put the interesting idea of QRFs in a new
language which may highlight its merits {(and perhaps, shortcomings)}, and may suggest directions
for further investigation.

 As mentioned above, our first goal is to eliminate the
use of coordinate systems in the {\it definition} of QRFs. 
In fact, 
making non-uniquely-determined  coordinate systems  play
important physical roles seems common  to the recent {\it concrete} applications of QRFs 
to the gravitational scenario of a mass in a superposition 
of position states~\cite{CGBB20,FB22,KHCB22,HKC23}: thus treating coordinate systems as
 if they  were
synonymous with 
``points of view'' or ``perspectives'' with respect to which
{\it observables} of
physical system are described---which they are not.\footnote{{{\label{perspneut}}
 At a late stage of development of this paper, the elegant group-theoretic approach to QRFs called {\it perspective-neutral} (PN)
 came to our attention---see, e.g., Ref.~\cite{HGHLM21} for
 a treatment dealing with 
 arbitrary unimodular symmetry groups. While we build upon 
 the
 very
 concrete coordinate-based QRF implementation of Refs.~\cite{CGBB20,FB22,KHCB22,HKC23}---in a sense, making
 ours a ``bottom-up'' approach---the PN approach  starts
 at a higher level of abstraction and generality (in terms of
the physical systems it intends to describe) and adds assumptions
 only to the extent they are needed to show desirable 
 properties---in a typical ``top-down'' approach.
 Although there certainly are similarities between the PN approach and our own---especially as regards their goals of providing a {\it generally-covariant} formalism---their different group-
 and geometric-oriented frameworks, respectively, make a  comparison of them an interesting 
 subject more appropriate for  future work.  
 %\sout{But noteworthy exceptions include approaches whose key idea is that a QRF is a subsystem of the total system, together with a state of it with respect to which the remaining subsystems get described (in broadly ``relational'' terms). These `perspective-neutral' approaches have been richly developed using elegant group-theoretic ideas, cf.\  Refs.~\cite{HG20,HGHLM21}; and recently, using tetrad fields, cf.\  Ref.~\cite{JG24}. But these approaches’ relation to  our approach, though interesting, is unclear; and must be postponed to future work.}
 }}
 To make this point clear: although coordinate systems are
enough to determine components of tensors which 
characterize the physical system, they are {\it not} enough
(and, in fact, not necessary either) 
to characterize  observables 
of the same system, since,
in general, components have no direct  physical meaning.\footnote{ 
For a concrete example of the dangers of attributing direct physical meaning to
tensor components, see, e.g., Refs.~\cite{C94,VMC96}.}
If observables are to play an important role, then  the notion of
{\it observers} or {\it reference frames}---which are rigorously formalized
as {\it tetrads} or {\it tetrad fields}---must be introduced and we stress that
the freedom to do so  in a given spacetime is {\it not} the same
as the freedom to choose coordinate systems  in that same spacetime---the latter
being immensely larger. Moreover,
once observers or reference frames
are introduced,
nothing physical can depend on the choice of coordinates.

But having made this statement of intent, we should stress again (so as to avoid unrealistic expectations
 and being unfair to the existing literature) that our purpose here is merely to recast the QRF idea in geometric terms. We do {\it not}
claim that analyses carried out using our approach could not be  performed
using the
coordinate-based one. In fact,  since the freedom in choosing coordinate 
systems { in any given spacetime} is
larger than that in choosing ``points of view'' (i.e., tetrad fields)
in that same spacetime, it is natural to expect that by a judicious
use of coordinate systems { (e.g.,
anchoring its definition on some physical criteria)}
one could reach  the same conclusions as the ones drawn using
the
geometric approach. In an analogy, this would correspond to formulating (and 
working with) GR using only multivariable calculus on
${\mathbb R}^n$, without ever mentioning manifolds, tangent spaces, tensors as multilinear mappings, connections, and the
whole {\it abstract} framework of differential geometry. So, if there is a merit 
in our approach, it is to make a clear distinction between 
the choices which {\it can} have physical 
consequences and the ones which cannot---something which we believe gets blurred in
the coordinate-based
approach.

Our geometric formulation of QRFs will be based on two main ideas. First, we think of a QRF as encoding quantum uncertainty about what is the observer’s 
or apparatus's perception of time and space in a given spacetime. 
In Section~\ref{sec:QRFrestricted}, we characterize the apparatus's frame by a tetrad: i.e.\ in physical terms, by an infinitesimal clock and three infinitesimal orthogonal rulers, located at a spacetime point. So a QRF is a (suitably normalizable) ``wave-function” assigning a complex amplitude to such tetrads. Though simple, this formulation is sufficient to reproduce one of the key themes of the literature on QRFs: that a superposition in the state of the measured system can be ``transferred”, by a transformation between QRFs,  to a superposition of QRFs. (This is shown at the end of Section \ref{sec:QRFrestricted}.)\footnote{At a late stage in developing this paper, we learned of another formulation of QRFs that invokes tetrads, viz.\ Ref.~\cite{JG24}. In 
Section~\ref{sec:discussion}, we will briefly compare it with our formulation.}

Second: in Section \ref{sec:genQRF}, we invoke the result that an assignment of a {\it generic} basis of the tangent space, at each point of a given manifold, {\it defines} a metric on the manifold---simply by stipulating that at each point, the assigned basis is to be orthonormal, i.e.\ a tetrad, for the metric to be defined.
As a result, we can interpret a section of the frame bundle (i.e.\ the bundle over spacetime whose fibre over each point is the set of all bases of the tangent space at that point) as a specification of:\\
\indent \indent   (i) a metric field, i.e.\ a spacetime, since the basis within the section that lies above the point $p$ is stipulated to be orthonormal in that spacetime's geometry; together with: \\
\indent \indent  (ii) a tetrad field, viz.\ the very elements of the section, on that spacetime. \\
Accordingly, a {\it pair} of spacetimes, for instance (each with its own tetrad field), is specified by a pair of sections. Thus, 
now invoking the first idea above: by assigning complex amplitudes to  generic bases at {\it each} point, one can describe, in a local manner,
not only a superposition of a pair of spacetimes (when only two bases at each point
are singled out), but also an {\it arbitrary} superposition of geometries. 

So much for the two main ideas. Then, in the last main Section (Sec.~\ref{sec:beyond}) we apply this fibre bundle formulation of QRFs to the ``simple" scenario analyzed in the literature, which we  recalled above: viz.\ a test particle probing the gravitational field sourced by a macroscopic superposition of position states of a large mass. Section~\ref{sec:discussion} is a brief discussion Section. In particular, it compares our  formulation of QRFs with {\it another} fibre bundle framework for QRFs, viz.\ that of  Ref.~\cite{KHACGBB24}. Section \ref{sec:discussion} also gives a brief outlook on possible lines of  further inquiry. 

It will also help to have a prospectus of the paper, that does not use the jargon of fibre bundles. In these terms, the paper is organized as follows. In 
Section \ref{sec:QRFrestricted}, we introduce the notion 
of QRFs 
in a given, well-defined spacetime (which we call {\it restricted} in 
such a 
context). In Section~\ref{sec:genQRF}, by demanding that the notion of QRFs
be diffeomorphism
invariant,  we relate QRFs in
isometric spacetimes and, eventually, extend their definition to
generic superpositions of geometries. This extension prompts the discussion of Section~\ref{sec:kd}, emphasizing the
different roles that our mathematical description of QRFs can play: one of merely describing (superpositions of) points of view in (superpositions of) arbitrary
geometries---the {\it perspectival} conception---and another of describing the arbitrary superpositions of geometries in themselves---the {\it basic} conception.
We make it clear that in this work we focus on the perspectival conception of QRFs.
In Section \ref{sec:beyond}, we revisit the scenario which has been used
in the recent literature to motivate the introduction of QRFs. We argue that 
if one intends to apply the QRF idea beyond the linear-gravity 
regime---which has been the goal in the  recent 
literature---then
one must face the subtlety of the non-uniqueness of mapping 
between
diffeomorphically-related spacetimes.  In particular, this mapping may be
related to the {\it dynamics} of the
evolution of the source mass
to 
the superposition of position states---a dynamics whose treatment  lies beyond current abilities. Finally, in Section \ref{sec:discussion} we present our
final remarks. We also sketch the general idea behind our construction in
Fig.~\ref{fig:QRF}---which the reader is invited to look at even
before proceeding to the next Section, and to revisit while reading. 

Finally, note that throughout the paper, we use units in which
$c = 1$ and the abstract-index notation for tensorial quantities (see, e.g.,
Ref.~\cite{Wald84}). According to this notation, Greek letters are used 
for ``concrete'' indices---i.e., those which assume numerical values---from 
$0$ to $3$. Latin letters $i,j,...$ are used for ``concrete'' indices 
from $1$ to $3$. Finally, Latin letters from the beginning of the alphabet, 
$a,b,...$ (up to $h$), are used for ``abstract'' indices---i.e., indices which
do {\it not} assume numerical values but instead indicate, by their position and number,
the type of tensor which
the indexed object is.

\section{Quantum reference frames on a given spacetime:
superposition of perspectives
}
\label{sec:QRFrestricted}

In order to motivate our construction by step-by-step reasoning, let us 
begin by considering a generic spacetime $({\cal M}, g_{ab})$, where 
$g_{ab}$ is the metric tensor 
defined on a differentiable
manifold ${\cal M}$. Although 
this gives complete information about the stage on which all 
systems
evolve, our fragmented perception (and measurements) 
of space and time as separate entities
usually leads us to introduce the notion of a {\it reference frame} (RF), 
which basically tells how {\it absolute} (infinitesimal) 
spacetime  intervals are to be decomposed
into {\it relative} 
(infinitesimal) space distances and time lapses in the 
neighborhood of each event. Obviously, each decomposition 
only
makes sense for specific observers/apparatuses and is only necessary
when talking about {\it observables}---i.e., 
outcomes of measurements performed by these specific
observers/apparatuses.

Geometrically, this classical 
notion of a reference frame in region
${\cal O}\subseteq {\cal M}$ is implemented
by means of an 
assignment (which we shall take to be smooth) ${\cal M} \supseteq {\cal O} \ni p \mapsto
\{{\bf e}_\mu^a(p)\}_{\mu = 0,1,2,3}$, 
which to every event $p$ in ${\cal O}\subseteq 
{\cal M}$
assigns a tetrad
$\{{\bf e}_\mu^a(p)\}_{\mu = 0,1,2,3}$. 
Here, using abstract-index notation: $a$ simply says that each element of 
$\{{\bf e}_\mu^a(p)\}_{\mu = 0,1,2,3}$ is a tensor of type $(1,0)$---i.e., a 4-vector---and $\mu$ labels the four elements of this set of 
4-vectors.\footnote{This concrete index $\mu$  in ${\bf e}_\mu^a$ 
must not be confused with the concrete indices which label components of 
tensors. Some references avoid this confusion by using Latin capital letters
in the former case, writing, e.g., ${\bf e}_A^a$. On the other hand, Latin capital
letters are also often used to represent algebra-valued quantities. So, here, 
we stick to the simpler index notation which only distinguishes between
concrete and abstract indices.} So, the fact that
$\{{\bf e}_\mu^a(p)\}_{\mu = 0,1,2,3}$ is a tetrad means that  
$g_{ab}{\bf e}_\mu^a(p){\bf e}_\nu^b(p)
= \eta_{\mu\nu } :={\rm diag}(-1,1,1,1)$. At
$p$, 
${\bf e}_0^a(p)$ sets {\it both} direction {\it and} 
unit for what is to be considered as ``time''---i.e., ${\bf e}_0^a(p)$ {\it is}
the 4-velocity of the observer/apparatus which the tetrad is supposed to describe at
$p$---while
$\{{\bf e}_j^a(p)\}_{j= 1,2,3}$ then gives the space directions and 
units in such a way that $c =1$ is ensured.

Despite the simplicity of the notion of tetrad, it can be argued that it is adequate to encode whatever the observer’s apparatus happens to be, essentially because all measurements reduce, at bottom, to measurements of lengths and time-intervals; cf.\ for example Ref.~\cite{MPSV23}. For whatever distinguishes different apparatuses carried by the {\it same observer}---hence,
associated to the same tetrad---is modelled by the way the tetrad couples to
the system being observed so as to give the {\it observable} quantity. 
For instance, an apparatus described 
by the tetrad
$\{{\bf e}_\mu^a\}_{\mu = 0,1,2,3}$ which measures electric
field, along the spatial direction ${\bf e}_j^a$, is modeled by the
coupling 
$F_{ab}{\bf e}^a_0 {\bf e}^b_j$, where $F_{ab}$ is the Faraday tensor 
which encodes all the information about the electromagnetic field of the system
being observed. Were we interested in measuring the energy density, then the measuring procedure would be modeled by the coupling $T_{ab}{\bf e}^a_0 {\bf e}^b_0$, with $T_{ab}$ being the stress-energy-momentum tensor of the system. Note an important fact:
given the system and the observer/apparatus, the {\it observables}---$E_j :=F_{ab}{\bf e}^a_0 {\bf e}^b_j$ and $\rho:=T_{ab}{\bf e}^a_0 {\bf e}^b_0$ in the examples above---must be {\it scalars}, in the sense of being insensitive to choice of coordinate systems; (indeed, coordinates have not even been mentioned thus far).

In the more
elegant and concise language of fibre bundles, one can 
think of a reference frame as a smooth local
section of
the bundle ${\cal F}_{\rm o}[g_{ab}]$ 
of (pseudo-)orthonormal frames of $({\cal M},g_{ab})$. (So here, `o' stands for orthonormal.) This is the language
we shall adopt.

As a first attempt at implementing the 
idea of a QRF, we want to
introduce 
uncertainties related to
the state of motion of
observers/apparatuses
at each event. This could be easily implemented by defining a
probability measure on each fibre of ${\cal F}_{\rm o}[g_{ab}]$---which represents
the different choices of tetrad at $p$. 

However,
since we ultimately want 
this uncertainty to be quantum   in nature---with all its linear, complex structure---we shall here {\it define}  a {\bf QRF} to be
an  
assignment 
$\Psi:{\cal F}_{\rm o}[g_{ab}] \to {\mathbb C}$, which to every 
tetrad $\{{\bf e}_\mu^a(p)\}_{\mu = 0,1,2,3}$
at $p\in{\cal M}$ assigns a complex number 
(a ``probability amplitude'')
$\Psi(p;\{{\bf e}_\mu^a(p)\})$, such that
\begin{eqnarray}
 \|\Psi(p)\|^2:=\int_{O(3,1)}d\mu_L(\Lambda) \left|\Psi(p;\{\Lambda^\nu_\mu{\bf e}^a_\nu(p)\})\right|^2 <+\infty
\label{normp}
\end{eqnarray}
for {\it each} 
$p$ and  {\it any fixed} $\{{\bf e}_\mu^a(p)\}_{\mu = 0,1,2,3}$;
where $\Lambda\in O(3,1)$ (the Lorentz group; i.e., 
the entries $\Lambda^\nu_\mu$ of matrix $\Lambda$ satisfy
$\Lambda^\alpha_\mu\Lambda^\beta_\nu \eta_{\alpha\beta} = \eta_{\mu\nu}$)
and $\mu_L$ is the (unique up to 
a multiplicative factor) invariant Haar 
measure defined on $O(3,1)$. (Note that Eq.~(\ref{normp})  
{\it defines} the symbol
$\|\Psi(p)\|^2$, {\it not} an object $\Psi(p)$; in fact, we will have no use for 
the latter here.)

The definition Eq.~(\ref{normp}) is basic to the rest of this paper; and in the rest of this Section, we develop ideas based on it. First, we make five
general comments, numbered (1), (2) etc.; and then we give some examples (beginning with Eq.~(\ref{EvalueQRF}) below) and some discussion of transformations. 

(1): As the first general comment, note that since
$O(3,1)$ is a noncompact group,
this integrability condition
%, for each $p\in {\cal M}$,
implies that any particular choice of a QRF $\Psi$ necessarily
breaks Lorentz invariance:
which is expected when adopting a ``point of view.''
But note that
the result of the integration in Eq.~(\ref{normp})
is independent of the choice of 
$\{{\bf e}^a_\mu(p)\}$ held fixed. (From now on, we will write simply 
$\{{\bf e}_\mu^a\}$ instead of $\{{\bf e}^a_\mu(p)\}$ 
where it is understood that it stands for 
the tetrad at
an event---which should be clear by context. Moreover, 
when figuring in arguments of functions, we simplify
it further and write merely $\{{\bf e}\}$, so as not to
proliferate unnecessary indices---unless needed 
by context.)

(2): We shall refer to this
notion of QRF as {\it restricted}, and write  `rQRF'; since we will later generalise it. The reason for the generalisation will be clear in  Sec.~\ref{sec:genQRF}.
But in short, we will treat superpositions of geometries by using the idea that given one geometry, a non-orthonormal basis at a point $p$ is a tetrad, i.e.\ is orthonormal, for another geometry.

(3): The concept 
of rQRF introduces uncertainties on measurements of
space distances and time intervals (and, from them, in 
measurements of any other observable, cf.\ Ref.~\cite{MPSV23}) 
without, however, making the metric
tensor uncertain. That is: it allows for ``superpositions'' of ``points of view''---e.g., 
due to the possible 
states of motion 
of the observers/apparatuses at each event---on a
well defined spacetime geometry.
Note, also,  that
one can consider linear combinations of rQRFs 
in the obvious (point-wise in 
${\cal F}_{\rm o}[g_{ab}]$) way. They 
are like collections of
wave functions, each of  them being 
defined on the fibres $\pi_{\rm o}^{-1}(p)$ of 
${\cal F}_{\rm o}[g_{ab}]$; where 
$\pi_{\rm o}: {\cal F}_{\rm o}[g_{ab}]
\to {\cal M}$ is the canonical projection of the bundle.

(4): In addition to Eq.~(\ref{normp}), one might impose further (reasonable)
constraints, such as that $({\cal M},g_{ab})$ is time orientable 
and that
$\Psi$ has support
on tetrads whose ${\bf e}_0^a(p)$ is future 
directed---although violation of this latter condition 
may find application in indefinite causal 
order analyses~\cite{OCB12}.
Moreover, if $\Psi$ is meant to represent
(superposition of) ``points of view'' of {\it physical} 
observers/apparatuses, it seems reasonable that it must satisfy
some sort of transport (``Boltzmann-like'') differential equation, describing ``diffusion" 
in ${\cal F}_{\rm o}[g_{ab}]$---in addition, 
perhaps, to some
other ``dynamical'' equation---constraining its local behavior, particularly for nearby
events $p,q$ related by curves having  the vectors ${\bf e}_0^a(p)$ 
as tangents 
(for $\Psi(p;\{{\bf e}\})\neq 0$).
But here, we  
leave the discussion of these further constraints
aside, since they are not essential to this paper's
presentation of the general ideas. In other words, here we shall not 
impose further restrictions
on the space of {\it kinematical} QRFs which might eventually
lead to the space of
{\it physical}  QRFs.

(5): Note 
that we are not imposing integrability
of $\|\Psi(p)\|^2$ on ${\cal M}$ (with respect to the preferred 4-volume
element $\epsilon$, or $\epsilon_{abcd}$ in 
abstract-index notation, selected by the metric $g_{ab}$), since this can
be too restrictive. For instance, it would exclude 
most globally defined (Q)RFs,
common even in the classical context to represent  
families of observers covering the whole spacetime.
And even though such an integrability condition for $\|\Psi(p)\|^2$
may be requested, if convenient---hence obtaining a finite
$\|\Psi\|^2:=\int_{\cal M}\epsilon\,\|\Psi(p)\|^2$, perhaps representing
a (spacetime-)localized ``observation''---one must be careful about interpreting 
$\|\Psi(p)\|^2/\|\Psi\|^2$ as any sort of
``probability density'' on ${\cal M}$. 
In this case, it seems more reasonable to codify in 
$\|\Psi(p)\|^2$ a similar role as the one played by the
smearing 
%(compactly supported, smooth) 
functions 
$f\in C_0^\infty({\cal M})$ which are
used to define 
observables 
$\hat{A}[f] = \int_{\cal M}\epsilon\, f(p) \hat{A}(p)$
in quantum theory, since, in general, point-wise observables $\hat{A}(p)$ 
only exist as
operator-valued {\it distributions}. We will return to this issue, including smearing, in Subsection A below.

{So much by way of general comments on Eq.~(\ref{normp}).} In order to illustrate 
the effect of rQRFs, first on classical observables, let us consider a
specific example: the electric field
observed/measured at an event $p$,
w.r.t.\ a given rQRF $\Psi$. Let us assume that the electromagnetic field is completely described, at $p$, by the 
Faraday tensor $F_{ab}$. If a classical observer (or particle) were passing through $p$ 
with 4-velocity $u^a$, the electric field
he/she would measure (or be subject to) along the spatial direction characterized by the 
unit vector $n^a$ (with $g_{ab}u^a n^b = 0$) 
would be 
$E_n=F_{ab}n^au^b$. So, it is only natural to 
define the expectation value for the $j$ component of the electric field
in the rQRF $\Psi$ by
\begin{widetext}
\begin{eqnarray}
\langle E_j\rangle :=\|\Psi(p)\|^{-2} 
\int_{O(3,1)}d\mu_L(\Lambda) F_{ab}{\bf e}_\alpha^a{\bf e}_\beta^b
\Lambda_j^\alpha \Lambda_0^\beta
\left|\Psi(p;\{\Lambda{\bf e}\})\right|^2 ,
    \label{EvalueQRF}
\end{eqnarray}
\end{widetext}
where $\{{\bf e}_\mu^a\}$ is any (fixed)  tetrad at $p$ and
$\{\Lambda {\bf e}\}$ stands for $\{\Lambda^\nu_\mu {\bf e}_\nu^a\}$
when in
arguments of functions (following the convention set above).

Similarly, {\it any} classical 
observable $A$ can be expressed as a {\it scalar} 
(in the sense of being {\it coordinate-independent}) quantity 
constructed out of the
tensors characterizing the system {\it and} 
 the coupling between the system and the tetrad characterizing the 
observer/apparatus performing the measurement. Hence, by averaging
over the  tetrads with the probability distribution given by
$\left|\Psi(p;\{{\bf e}\})\right|^2$, one 
obtains the expectation value $\langle A\rangle$ at 
$p$. 

A similar rationale  applies to quantum observables.  Any 
quantum observable can be represented by a {\it scalar} operator-valued
distribution
$\hat{A}$ which depends on the physical system and on the 
observer/apparatus 
performing the measurement (i.e., on the tetrad which characterizes the latter and its coupling to the system). Let us consider, again, a concrete example: the operator-valued
distribution
$\hat{\cal \rho} = 
{\bf e}_0^a {\bf e}_0^b \hat{T}_{ab}$
which describes 
the  energy density, with respect to a given observer/apparatus with
 4-velocity
${\bf e}_0^a$,
of a system whose (renormalized) energy-momentum tensor is described by the
operator-valued distribution
$\hat{T}_{ab}$.
Thus, if the system is in a state $|s\rangle$---at this point supposed to be an observer-independent  statement---then 
the expectation value of its energy density
w.r.t.\ $\Psi$
is given by (recalling that $\epsilon$ is 
the preferred 4-volume element on $({\cal M},g_{ab})$
and using $\|\Psi\|^{-2}=1/\|\Psi\|^2
=1/\int_{\cal M}\epsilon \|\Psi(p)\|^2$) 
\begin{widetext}
\begin{eqnarray}
\langle {\cal \rho}\rangle :=
\|\Psi\|^{-2}
\int_{\cal M}\epsilon\,
\int_{O(3,1)}d\mu_L(\Lambda) 
\left|\Psi(p;\{\Lambda{\bf e}\})
\right|^2\Lambda^\alpha_0
\Lambda^\beta_0
{\bf e}_\alpha^a
{\bf e}_\beta^b
\langle s|\hat{T}_{ab}|s\rangle 
,
    \label{EvalueQRFrho}
\end{eqnarray}    
\end{widetext}
again for any fixed tetrad field
$\{{\bf e}_\mu^a\}$. Note that, now, we  considered
a square-integrable (in fact, a compactly-supported)
$\|\Psi(p)\|^2$, due to the distributional nature of quantum
observables. 

Now, let us consider changing the rQRF. 
Different rQRFs can lead to different
values for the observable quantities, since they describe 
different
``superpositions of perspectives.''
An exception to this may occur when the
rQRFs are related
by an 
{\it isometry} of $({\cal M}, g_{ab})$, $\iota:{\cal M}\to {\cal M}$, 
via\footnote{\label{fn:phase}It is worth keeping in mind that the use of QRFs proposed to calculate expectation values, 
Eqs.~(\ref{EvalueQRF}) and (\ref{EvalueQRFrho}), allows for an arbitrary phase function $\varphi(p;\{{\bf e}\})$
to be included in
Eq.~(\ref{eq:iota}): 
$\widetilde{\Psi}(p;\{{\bf e}\}):= e^{i\varphi(p;\{{\bf e}\})}
\Psi(\iota^{-1}(p);\{\iota_\ast{\bf e}\})$. This could be useful in the analysis
of Subsec.~\ref{subsec:flat} below.
}
\begin{eqnarray}
\Psi\mapsto
\widetilde{\Psi}(p;\{{\bf e}\}):=
\Psi(\iota^{-1}(p);\{\iota_\ast{\bf e}\}),    
\label{eq:iota}
\end{eqnarray}
where $\iota_\ast$ is the associated pull-back mapping 
between tangent vectors at $p$ and at $\iota^{-1}(p)$. 
It 
follows directly from the definition
that any observable of a system in a configuration/state
{\it which respects this isometry}---in particular, 
any geometric observable---will be invariant under this particular
transformation.

Note, also, that one can 
 consider linear combinations 
of $\Psi$ and $\widetilde{\Psi}$ to get other rQRFs on
$({\cal M}, g_{ab})$.  So, the set of rQRFs on a {\it given}
spacetime is a complex vector space where the isometries of 
the spacetime (if there are any) can be
naturally represented.

Although our definition of QRFs has relied totally on geometric
objects, it may be convenient
to introduce coordinate systems (CSs) when carrying out explicit calculations.
Given a rQRF $\Psi$ and a CS 
$\chi:{\cal O} \to U \subseteq {\mathbb R}^4$
defined on ${\cal O}\subseteq {\cal M}$,
we obtain a {\it representation} $[\Psi]_\chi$ of $\Psi$, by requiring that, at the event with coordinates $x^\mu$,
and at
the tetrad with {\it  components} $e^\alpha_{(\beta)}$  in the coordinate 
basis $\{\partial_\alpha^a\}$ 
induced by $\chi$---i.e.,
${\bf e}_\beta^a = e_{(\beta)}^\alpha\partial_\alpha^a$---$[\Psi]_\chi$ is to assume the value that $\Psi$ takes there. (The use of parentheses
in the lower index of $e_{(\beta)}^\alpha$ is merely to differentiate
the labeling of each tetrad element from the labeling of the components of these
elements.) Thus we define:
\begin{eqnarray}
[\Psi]_\chi(x^\mu; \{e^\alpha_{(\beta)}\}):=\Psi(\chi^{-1}(x^\mu);
\{e^\alpha_{(\beta)}\partial_\alpha^a\}).
    \label{Psicoord}
\end{eqnarray}
Note that  the indices in this definition 
belong to the arguments of the functions, not to the functions 
themselves, which are {\it scalars}---hence, there is no need for matching indices on the two sides.

It is important to note that
changing the coordinate system only changes the 
representation $[\Psi]_\chi$, but {\it not}  $\Psi$
itself---hence, it  has no effect on 
expectation values of observables.
These two
concepts, reference frames and coordinate systems, are
often confused with one another, even in classical physics.
As we have explained before, the former prescribes how
spacetime is to be described, by observers/apparatuses, 
as space and time separately. So it {\it can} have effects on the value of observables/measurements performed
on a given physical system; (although of course, the physical system, itself, is
oblivious to the adoption, or not, of a reference frame---at least in classical physics).
The latter, on the other hand, is a mere (arbitrary) numerical labelling 
and, as such,
has no consequences for physical results. In the context above, this means that given a rQRF
$\Psi$, no physical observable can depend on which representation
$[\Psi]_\chi$ is used to perform the calculations. Changes in the coordinate 
system (to be interpreted as {\it passive} diffeomorphisms) lead to
different 
representations which, nonetheless, {\it must} lead to 
the same physical observations.

Before ending this Section, it is 
important to emphasize, once more,  that, as far as the spacetime 
geometry is concerned, different rQRFs 
only represent different
``perspectives'' of the {\it same} physical situation. 
In the same way 
that a tetrad $\{{\bf e}_\mu^a\}$ can stand for an idealization of
a classical observer/apparatus performing time and distance measurements (from which, strictly speaking, any other measurement is obtained~\cite{MPSV23}),
a rQRF $\Psi$ can be thought of as an idealization (perhaps overly
generalized) of 
observers/apparatuses
which
can exhibit quantum aspects, such as spacetime delocalization and 
velocity uncertainty.

One particularly simple (and idealized)
example of a rQRF,
defined with the help of a coordinate system, is one completely 
concentrated on a time-like  curve described by
$x^\mu(\tau)$---representing the perspective of
a point-like 
observer/apparatus or test particle having the curve as its worldline:
\begin{eqnarray}
supp[\Psi]_\chi = \left\{ \left(x^\mu(\tau),\{u^\alpha(\tau),e^\alpha_{(j)}(\tau)\}\right); \;\tau\in I\subseteq
{\mathbb R}
\right\},
    \label{rQRFcurve}
\end{eqnarray}
with $u^\alpha(\tau) = dx^\alpha(\tau)/d\tau$ being the components of its 4-velocity
and $\tau$ its proper time ($e^\alpha_{(j)}(\tau)$ are the components of the
other elements of the tetrad, which are left unspecified here).
From this, less trivial examples---which play a significant role, e.g., 
in Ref.~\cite{HKC23}---can be constructed by superposition,
possibly representing  ``probes'' whose location may not be well defined.
We shall come back to this later.

\subsection{An application to flat spacetime}
\label{subsec:flat}

Before considering more general scenarios (involving
possible superposition of different geometries),
let us illustrate
this geometric approach in  flat spacetime.
In flat spacetime, there are preferred families of inertial 
observers, namely, the ones
characterized by {\it uniform} tetrad fields. 
It is common to use these
families to interpret/characterize the states of the 
system. 
Here, insisting on the 
use of tetrads may seem pedantic, since there is for each  of
these families a natural choice of coordinates
which faithfully characterizes time and space measurements, viz.\ the usual inertial Cartesian coordinates. 
However,  our purpose here is to illustrate the 
use of the tetrad-based notion of QRF---and this faithful 
characterization in terms of  inertial Cartesian coordinates is
restricted to {\it this} class of inertial families of observers in {\it flat}
spacetime.

First, if we want to consider superposing different ``points of view,''  we have to make explicit the dependence of observables on
observers. Indeed, there are two aspects to this. First, we recall the need, familiar in quantum field theory, to smear observables with test functions $f$ defined on the spacetime $({\cal M},g_{ab})$. (But what follows will not depend on the details of quantum field theory.) 
This is done, as usual, by having for a given observable $A$, a smearing map,  $C^\infty_0({\cal M}) \ni f \mapsto A_f\in {\cal L}({\cal H})$,  which takes
compactly supported, smooth functions $f$ on  $\cal M$ to linear operators $A_f$ acting on the Hilbert space ${\cal H}$.

But now, we   
extend the idea of this definition, so that the domain is the compactly supported, smooth functions $F$ 
defined on the 
 entire bundle of (pseudo-)orthonormal frames, 
 ${\cal F}_{\rm o}[g_{ab}]$. This extension is not {\it ad hoc}, but accords 
  with our philosophy of  talking about observables only once observers are explicitly mentioned.

Thus we have a smearing function 
$\hat{A}:C^\infty_0({\cal F}_{\rm o}[g_{ab}]) \to {\cal L}({\cal H})$; by which 
compactly supported, smooth functions $F$ 
defined on ${\cal F}_{\rm o}[g_{ab}]$ are (linearly) mapped to 
operators $\hat{A}[F]$. Formally, we write:
\begin{eqnarray}
    \hat{A}[F] = \int_{\cal 
    M}\epsilon\int_{O(3,1)}d\mu_L(\Lambda)
    F(p;\{\Lambda{\bf e}\}) \hat{A}(p;\{\Lambda{\bf e}\}).
    \label{Ape}
\end{eqnarray}
Note that in Eq.~(\ref{Ape}), $\hat{A}(p,\{{\bf e}\})$ is an 
operator-valued
distribution---for instance,  in the example of 
Eq.~(\ref{EvalueQRFrho}),
$\hat{A}(p,\{{\bf e}\})
= \hat{T}_{ab}(p){\bf e}_0^a(p){\bf e}_0^b(p)$.

Consider, now, a quantum {\it mechanical} system and let $|s\rangle$ be its (observer-independent) 
state. (Here, we say `mechanical' because we shall not here consider the subtleties of quantum field theory, such as the non-localizability of its states and the non-existence of finite-rank spectral projectors.) An inertial family of observers, $O$, described by the {\it uniform} tetrad field 
$\{{\bf o}_\mu^a\}$, can characterize the system (in a compact region)  by the expectation values of  self-adjoint operators $\hat{A}_O[f]$ which, for them, represent observables $A_O(p)$ smeared with test functions $f\in{C_0^\infty}({\cal M})$.

We thus envisage a rQRF $\Psi_O$ which describes the {\it classical} family of observers 
$O$---i.e., ``peaked'' at $\{{\bf o}_\mu^a\}$---and
with $\|\Psi_O(p)\|^2 = f(p)$.
Thus we write:
\begin{eqnarray}
\hat{A}_O[f] &=& \int_{\cal M}\epsilon
f(p) \hat{A}_O(p)
\nonumber \\
&=& \int_{\cal M}\epsilon 
\int_{O(3,1)}
\!\!\!
d\mu_L(\Lambda)
\left|\Psi_O(p;\{\Lambda{\bf e}\})\right|^2 \hat{A}(p;\{\Lambda{\bf e}\}) 
\nonumber \\
&=& \hat{A}[|\Psi_O|^2].
    \label{eq:AOf}
\end{eqnarray}
Note that this equation is only defined for nonnegative test functions $f$,
since we set $f(p) = \|\Psi_O(p)\|^2$. Nonetheless, this is enough to define,
by linearity,
the mappings $\hat{A}$ on arbitrary functions since an arbitrary 
real 
function $F$ can be (nonuniquely) 
written as $F = F_+ - F_-$, with both $ F_+$ and $ F_-$ being nonnegative functions. (The decomposition becomes  unique if we further impose
$F_+ F_- = 0$ pointwise.) In what follows, we suppose that Eq.~(\ref{eq:AOf}) is also applicable for a generic QRF $\Psi$---as we have already done when  proposing Eq.~(\ref{EvalueQRFrho}).

To close this Section, we now suppose that  the
state $|s\rangle$  can be written as a coherent superposition,
$|s\rangle = \sum_{I\in{\cal I}}c_I|s_I\rangle$ (${\cal I}$ being some index set), with the particular property that
there are spacetime isometries
$\{\iota_I\}_{I\in {\cal I}}$ whose unitary 
representations $\{\hat{U}_I\}_{I\in{\cal I}}$ 
on ${\cal H}$ relate some fiducial
state, say  $|s_0\rangle$, to each state $|s_I\rangle$: $\hat{U}_I|s_0\rangle = |s_I\rangle$, $I\in{\cal I}$.

We will now show that our framework of QRFs, using maps $\Psi: {\cal F}_{\rm o}[g_{ab}] \rightarrow {\mathbb C}$, is able to ``transfer the superposition" in the state $|s\rangle = \sum_{I\in{\cal I}}c_I|s_I\rangle$ ``onto" the set of quantum reference frames that are related [as in 
Eq.~(\ref{eq:iota})] to a given QRF $\Psi$ by the isometries $\iota_I$. Here ``transferring a superposition” will amount to an equality of expectation values. That is: we will show that the expectation value of a quantity $A$ in a given QRF, i.e.\ smeared as in Eq.~(\ref{eq:AOf}),  for the state $|s\rangle = \sum_{I\in{\cal I}}c_I|{s}_I\rangle$,  can equal the expectation value of the  quantity $A$, as smeared according to some linear combination of isometrically defined QRFs, for the fiducial state  $|{s}_0\rangle$. 

To be vivid,  we can suppose, for example, that: (i)~according to the given QRF $\Psi$, the state $|{s}_0\rangle$ is well-localised spatially
(though not necessarily at any particular point which one chooses to call ``the origin''---by, e.g., assigning a particular value of  $\Psi$ at it);
 and (ii)~the isometries $\iota_I$ are spatial translations (not necessarily in the same spatial direction).   For such a case, we will show  that it is possible
 to have equality between: (a)~the expectation value of $A$ attributed by the QRF $\Psi$, as smeared by some (not necessarily well-localised) $\Psi$ 
 [using Eq.~(\ref{eq:AOf})], for the spatially superposed state $|s\rangle = \sum_{I\in{\cal I}}c_I|{s}_I\rangle$; and 
(b)~the expectation value of $A$ attributed by a ``weighted'' set of  QRFs $\widetilde{\Psi}_I$, $I\in {\cal I}$, each related to $\Psi$ by the spatial translation $\iota_I^{-1}$---i.e.,   smeared  instead by the corresponding linear combination of isometrically-defined QRFs $\widetilde{\Psi}_I$ [Eq.~(\ref{eq:iota}) with $\iota = \iota_I^{-1}$]---now for the single well-localised state $|{s}_0\rangle$. 

But we stress that we mention this spatial example only for vividness. The calculation below is valid  for any fiducial state $|s_0\rangle$ that is related by spacetime isometries $\{\iota_I\}_{I\in {\cal I}}$ to the states  $|s_I\rangle$ in the expansion of  $|s\rangle$ as $ \sum_{I\in{\cal I}}c_I|s_I\rangle$. 
Thus we calculate, in general:
\begin{widetext}
  \begin{eqnarray}
  \langle s| \hat{A}[|\Psi|^2]|s\rangle &=& { 
  \sum_{I,J\in{\cal I}}c_J^\ast c_I\langle s_J| \hat{A}[|\Psi|^2]|s_I\rangle}
%  \nonumber \\
% &=& 
 = \sum_{I\in{\cal I}}|c_I|^2\langle s_0|\hat{U}^{\dagger}_I 
  \hat{A}[|\Psi|^2]\hat{U}_I|s_0\rangle
%  \nonumber \\
% & & +
 + \sum_{I\neq J\in{\cal I}}c_J^\ast c_I\langle s_J|
  \hat{A}[|\Psi|^2]|s_I\rangle
  \nonumber \\
  &{=}& 
  { \sum_{I\in{\cal I}}|c_I|^2\langle s_0|
  \hat{A}[|\widetilde{\Psi}_I|^2]|s_0\rangle+
  \sum_{I\neq J\in{\cal I}}c_J^\ast c_I\langle s_J|
  \hat{A}[|\Psi|^2]|s_I\rangle
  }
  \nonumber \\
  &=& 
  \left< s_0\left|
  \hat{A}\left[\left|\sum_{I\in{\cal I}}c_I\widetilde{\Psi}_I\right|^2\right]\right|s_0\right>
% \nonumber \\
%  & & +
 + \sum_{I\neq J\in{\cal I}}c_J^\ast c_I\left[\langle s_J|
  \hat{A}[|\Psi|^2]|s_I\rangle-\langle s_0|
  \hat{A}[\widetilde{\Psi}^\ast_J\widetilde{\Psi}_I]|s_0\rangle\right],
        \label{eq:Mink1}
  \end{eqnarray}
  \end{widetext}
where (i) as anticipated, in accordance with Eq.~(\ref{eq:iota}) (with
$\iota = \iota_I^{-1}$), we have defined $\widetilde{\Psi}_I(p;\{{\bf e}\}):=
{\Psi}(\iota_I(p);\{\iota_I^\ast{\bf e}\})$ and (ii) we have
used the invariance of the measure/volume elements 
whenever needed to demonstrate the identity 
\begin{widetext}
\begin{eqnarray}
\langle s_0|\hat{U}^{\dagger}_I 
  \hat{A}[|\Psi|^2]\hat{U}_I|s_0\rangle
  &=&\left< { s}_0\left|\int_{\cal M}\epsilon
  \int_{O(3,1)}d\mu_L(\Lambda) 
  \left|\Psi(p;\{\Lambda {\bf e}\})\right|^2 \hat{A}(\iota_I^{-1}(p);
  \{\iota_{I\ast}(\Lambda{\bf e})\})
  \right|{ s}_0\right>
  \nonumber \\
  &=&\left< { s}_0\left|\int_{\cal M}\epsilon
  \int_{O(3,1)}d\mu_L(\Lambda) 
  \left|\Psi(p;\{\Lambda {\bf e}\})\right|^2 \hat{A}(\iota_I^{-1}(p);
  \{(\Lambda\Lambda_I^{-1}){\bf e}\})
  \right|{ s}_0\right>
  \nonumber \\
  &=&\left< { s}_0\left|\int_{\cal M}\epsilon
  \int_{O(3,1)}d\mu_L(\tilde{\Lambda}\Lambda_I) 
  |\Psi(\iota_I(p);\{(\tilde{\Lambda}\Lambda_I) {\bf e}\})|^2 \hat{A}(p;
  \{\tilde{\Lambda}{\bf e}\})
  \right|{s}_0\right>
  \nonumber \\
  &=&\left< s_0\left|\int_{\cal M}\epsilon
  \int_{O(3,1)}d\mu_L(\tilde{\Lambda}) 
  |\Psi(\iota_I(p);\{\iota_I^\ast(\tilde{\Lambda} {\bf e})\})|^2 \hat{A}(p;
  \{\tilde{\Lambda}{\bf e}\})
  \right|s_0\right>
   \nonumber \\
  &=&\left< { s}_0\left|\int_{\cal M}\epsilon
  \int_{O(3,1)}d\mu_L(\Lambda) 
  |\widetilde{\Psi}_I(p;\{\Lambda {\bf e}\})|^2 \hat{A}(p;
  \{{\Lambda}{\bf e}\})
  \right|{ s}_0\right>
  \nonumber \\
  & =& 
  \langle {s}_0|
  \hat{A}
  [|\widetilde{\Psi}_I|^2]
  |{ s}_0\rangle
    \label{eq:ident}
\end{eqnarray}
\end{widetext}
which we used passing from the first to the second line of Eq.~(\ref{eq:Mink1}).

The result shown in Eq.~(\ref{eq:Mink1}), whose validity is quite general, 
is very reasonable: it means that it is {\it possible} to transfer the coherent
superposition of states in ${\cal H}$ to a superposition of QRFs, in what
concerns the
observable $A$,
provided the term in square brackets in the last line vanishes---i.e.,
the ${\cal H}$-off-diagonal terms $\langle s_J|
  \hat{A}[|\Psi|^2]|s_I\rangle$ match the QRF-off-diagonal terms
  $\langle s_0|
  \hat{A}[\widetilde{\Psi}^\ast_J\widetilde{\Psi}_I]|s_0\rangle$. Obviously, this will
  not be true for an arbitrary QRF $\Psi$---although the freedom mentioned in footnote~\ref{fn:phase} can be put to some use here---and it will, in general, depend on the observable $A$. As we see matters, this dependence may be taken to point to: (i)~a limitation of our proposed implementation of QRFs (or of the generalization of Eq.~(\ref{eq:AOf}) to a generic $\Psi$); or (ii)~a feature of the QRF idea itself; or (iii)~the need to impose constraints on $\Psi$
(such as, e.g., the ones which would restrict $\Psi$ to belong to the ``physical'' QRF space---see the general comment (4) in Sec.~\ref{sec:QRFrestricted}). We postpone to future work the investigation of these three alternatives.  Here, we just note that, according to our understanding, a similar dependence on the observable ${A}$ is also present  in the coordinate-based formulation of QRFs. In applications of QRFs to  non-relativistic quantum mechanics, for instance, one has to choose whether  to privilege the position or the momentum representation before deciding which are the  relevant QRF transformations (see, e.g., Ref.~\cite{GCB19}).

\section{Generalized QRF\lowercase{s} and superposition 
of geometries}
\label{sec:genQRF}

In the context of dynamical theories for the metric field $g_{ab}$ (such as  GR), 
it is widely assumed that the field equations are
{\it diffeomorphism invariant}. This 
means that given {\it any}
 diffeomorphism $\phi:{\cal M}\to {\cal M}$,
the spacetimes $({\cal M},g_{ab})$ and  $({\cal M},\phi^\ast g_{ab})$
are {\it physically} indistinguishable---where $\phi^\ast$ is the 
push-forward mapping between tensors defined at $p$ and at
$\phi(p)$ (i.e., $\phi$ here is seen as an {\it active} diffeomorphism).
According to this view, points $p$ of the underlying 
manifold ${\cal M}$ have no absolute meaning by themselves, acquiring
significance only 
to the extent that they can be characterized by physical quantities
evaluated
at them. 
This view  is of course the legacy of Einstein's hole argument. For discussion, cf.\ e.g.\ Ref.~\cite{S14}; and for a recent perspective which is close to this paper's fibre bundle approach to QRFs, and which we further discuss in Section \ref{sec:discussion}, cf. Refs.~\cite{GB23,GB23a,G24}.\footnote{There are of course meanings of `general covariance’ that are logically stronger than diffeomorphism invariance. We will not need to discuss them. But we note, for example, that Anderson in  Ref.~\cite{A67} argued that GR is distinctive in that all fields, even the metric and connection, are dynamical, rather than ``absolute” or a ``fixed canvas": so that the traditional requirement that the symmetry group of a theory formulated on a manifold $\cal M$  should preserve absolute geometric objects is vacuously satisfied by {\it any} diffeomorphism of the manifold $\cal M$ just because there are no such absolute objects, i.e.\ Diff($\cal M$)  is the symmetry group of GR (taken as using only one manifold $\cal M$). But as subsequent literature showed, it is hard to define `absolute'  precisely so as to get the intuitively right verdicts for all theories; cf.\ e.g.\ Ref.~\cite{P06}.}

Now, let us analyze how this invariance 
of the classical theory
fits into the scope of QRFs.
It is {\it not} true that any given diffeomorphism 
$\phi$
induces a 
transformation between rQRFs 
on $({\cal M}, g_{ab})$ via 
Eq.~(\ref{eq:iota}) (with $\iota$ replaced by $\phi$). For
given 
a tetrad $\{{\bf e}^a_\mu\}$ at $p$,
in general $\{\phi_\ast {\bf e}^a_\mu\}$ will fail to be a tetrad,
according to $g_{ab}$ at $\phi^{-1}(p)$ (unless $\phi$ 
is an isometry). 

Notwithstanding this, 
we can consider that
$\phi$ induces 
transformations {\it between} rQRFs defined on 
$({\cal M}, g_{ab})$ {\it and} 
rQRFs defined on $({\cal M}, \phi^\ast g_{ab})$, 
since these spacetimes are,
by their very definition, isometric. 
However, now, a superposition
of $\Psi$ and $\widetilde{\Psi}$ is no longer technically possible, since 
their domains
(the bundles of orthonormal frames ${\cal F}_{\rm o}[g_{ab}]$ and
${\cal F}_{\rm o}[\phi^\ast g_{ab}]$, respectively) 
are, in general, distinct.

In order to circumvent this difficulty, we extend 
the definition
of a QRF to be an assignment defined on the {\it bundle of frames}
${\cal F}({\cal M})$ on the manifold ${\cal M}$. Beware of ambiguity: the  `frame-bundle' or `bundle of frames'  ${\cal F}({\cal M})$ has as the fibre above a point $p \in {\cal M}$ the set of all (so: not necessarily (pseudo-)orthonormal) {\it bases} of $T_p$---{\it not} the set of all tetrads, i.e.\ `frames' in our physical sense. Hence our  use, since the start of Sec.~\ref{sec:QRFrestricted}, of the subscript `o' for `orthonormal' for the bundle  ${\cal F}_{\rm o}[g_{ab}]$, whose fibres {\it do} consist of tetrads according to the given $g_{ab}$.

 That is: we now define a QRF  to be a map,
$\Psi:{\cal F}({\cal M}) \to {\mathbb C}$,
that to every (not necessarily orthonormal in the aforementioned metric $g_{ab}$) {\it basis} $\{{\bf x}_\mu^a\}$ of the tangent space at
$p\in {\cal M}$ assigns a complex value 
$\Psi(p;\{{\bf x}\})$. \\
More precisely, we ``work locally'' in $\cal M$, and so define $\Psi$ on the bundle's points lying above some region ${\cal O} \subseteq {\cal M}$. That is: $\Psi$ is defined on some $\pi^{-1}(O)$, where $\pi:{\cal F}({\cal M})\to {\cal M}$ is the canonical projection of the frame bundle.

The reader may wonder why we went
through the trouble of restricting,
at first,
the definition of $\Psi$ to the bundle of orthonormal frames
(obtaining the rQRFs)
only to later extend it to the bundle of frames. Why not 
consider this extended  conception of a QRF from the beginning? 

The reason is for the sake of clarity. For although now $\Psi$ is defined for any basis
$\{{\bf x}^a_\mu\}$ at $p$, its {\it meaning}
continues to be the same as before, i.e., 
that of a ``complex amplitude'' assigned to
the basis $\{{\bf x}^a_\mu\}$ {\it seen as a tetrad} at 
$p$---and hence referring to
a spacetime whose metric $g_{ab}$
at $p$ satisfies $g_{ab}{\bf x}_\mu^a{\bf x}_\nu^b = \eta_{\mu \nu}$. 

Thus the idea is that {\it any} section of ${\cal F}({\cal M})$ (i.e., a basis field)  can be seen as
a section of ${\cal F}_{\rm o}[^{({\bf x})}\!g_{ab}]$ (i.e., a tetrad field) for 
{\it some} (uniquely defined) metric field $^{({\bf x})}\!g_{ab}$, namely
$$
^{({\bf x})}\!g_{ab} := \eta_{\mu \nu} {\bf X}_a^\mu {\bf X}_b^\nu,
$$
where $\{{\bf X}_a^\mu\}$ is the dual-basis field related 
to $\{{\bf x}^a_\mu\}$ (i.e.,
satisfying ${\bf X}_a^\mu {\bf x}^a_\nu = \delta^\mu_\nu$  at each manifold
point).\footnote{Here, again, the concrete index  $\mu$ in ${\bf X}^\mu_a$ labels the different basis elements (as in ${\bf e}_\mu^a$ and ${\bf x}_\mu^a$).}

Thus a spacetime with a specific metric is given uniquely 
by a section of $\cal F({\cal M})$, the bundle of all frames (although
different sections can lead to the same spacetime). {\it More precisely: a section specifies a spacetime, together with a tetrad field  (i.e., a point of view) 
on it, since the elements of the section are bases at the spacetime points,  that are orthonormal in that spacetime.}

 As a result of this, a generic QRF $\Psi:{\cal F}({\cal M}) \to {\mathbb C}$ can represent  {\it not only }
a superposition of ``points of view'' on a given classical
spacetime---if ${supp}(\Psi)\subset {\cal F}_{\rm o}
[g_{ab}]$ for some metric field $g_{ab}$---{\it but also}
a superposition of spacetime metrics in the region $\pi({supp}(\Psi))\subseteq {\cal M}$; 
(provided of course that these metrics are defined on 
the same base manifold ${\cal M}$).
  Thus the idea of stipulating a basis-field to be a tetrad-field is the device by which Sec.~\ref{sec:QRFrestricted}'s simple and restricted  conception of a QRF can be applied so as to describe a superposition of spacetime metrics.\footnote{\label{announce7person}{Our invoking the frame bundle prompts a comparison with a different  use of fibre bundles as a framework for QRFs, recently proposed by Ref.~\cite{KHACGBB24}; (cf.\
also Refs.~\cite{GB23, GB23a}, and Ref.~\cite{G24a} Sections 3, 7 and Appendix A). We will make this comparison in Sec.~\ref{sec:discussion}.}}

In Sec.~\ref{sec:beyond}, we will apply the above ideas to describe the gravitational field due to a macroscopic spatial superposition of a large mass. So to prepare for the details of that, we end this Section by stating how the normalisation condition, Eq.~(\ref{normp}), and the definition of a rQRF transformation by an isometry, Eq.~(\ref{eq:iota}), get generalised in the setting of generic QRFs.

Thus the normalisation condition~(\ref{normp}) is now replaced by
\begin{eqnarray}
 %   \item[(i)] 
 \|\Psi(p)\|^2:=\int_{GL(4)}d\mu_G(M) \left|\Psi(p;\{M{\bf x}\})\right|^2 <+\infty
\label{normg}
\end{eqnarray}
for all 
$p$ and {\it any fixed} $\{{\bf x}_\mu^a\}$,
where $M\in GL(4)$ (i.e., $M$ is an arbitrary, invertible
$4\times 4$ real matrix)
and $\mu_G$ is the (unique up 
to a multiplicative
factor)
invariant
Haar
measure defined on 
$GL(4)$.
(Note that, again, the result of the integration in Eq.~(\ref{normg}) 
does 
{\it not} depend on the choice of the fiducial $\{{\bf x}_\mu^a\}$ held fixed.)

Also, now, any diffeomorphism $\phi$ induces 
%a mapping in ${\cal F}$
a QRF transformation 
$\Psi \mapsto \widetilde{\Psi}$ 
 via [cf.\ Eq.~(\ref{eq:iota})]:
\begin{eqnarray}
\widetilde{\Psi}(p;\{{\bf x}\}) :=
\Psi(\phi^{-1}(p);\{\phi_\ast{\bf x}\}),
\label{PsionF}
\end{eqnarray}
which, in particular, maps rQRFs on 
$({\cal M},g_{ab})$ into 
rQRFs on $({\cal M},\phi^\ast g_{ab})$. It is important to stress, though,
that a generic
QRF {\it cannot} be interpreted as a  rQRF on some spacetime; for instance,
a linear complex combination of $\Psi$ and $\widetilde{\Psi}$ given above
(which is now possible)
is not, in general, a rQRF even if $\Psi$ and (consequently) $\widetilde{\Psi}$
are.

In this new framework,  
{\it classical} diffeomorphism invariance---i.e.\ the
physical equivalence of $({\cal M},g_{ab})$ and  
$({\cal M},\phi^\ast g_{ab})$---is ensured by
imposing invariance of all (geometric) observables under the map 
$\Psi\mapsto \widetilde{\Psi}$ given by Eq.~(\ref{PsionF}).

It is important to point out that, in contrast to rQRFs 
defined on isometric spacetimes---to 
which the second-to-last 
paragraph before Subsec.~\ref{subsec:flat} still applies---two 
generic QRFs, or even two rQRFs defined on nonisometric spacetimes,
describe completely 
different physical situations (not only different ``perspectives''
 on a given, classical spacetime). In fact, 
as pointed out above, since {\it any} spacetime
(with base manifold ${\cal M}$) can be (uniquely) characterized by
a section of ${\cal F}({\cal M})$---viz.\ it is that spacetime for which the given section
is also a section of
${\cal F}_{\rm o}[g_{ab}]$---a QRF may describe an
arbitrary superposition of spacetimes (with the same base manifold).
This generality  of QRFs in comparison to rQRFs prompts the discussion in the next Section.

\section{Perspectival \lowercase{vs.}\ Basic  QRF
conceptions}
\label{sec:kd}

Notice that when we defined a rQRF
in Sec.~\ref{sec:QRFrestricted},  say $\Psi_r:{\cal F}_{\rm o}[g_{ab}]\to
{\mathbb C}$, we
did not attribute to $\Psi_r$ the role of {\it determining} the geometry of the
spacetime. $\Psi_r$ merely represented possible superpositions of the 
``points of view'' (i.e., tetrads)
consistent with the {\it given} classical 
 geometry. As such, $\Psi_r$ did not need
to be defined globally---i.e.,  there  was no need for 
$\pi_{\rm o}(supp(\Psi_r)) = {\cal M}$. One can consider, for instance, a
$\Psi_r$ that
represents the perspective of a  particle in a superposition of worldlines
evolving in the given spacetime, in which case a natural choice for $\Psi_r$ would be supported on curves in ${\cal F}_{\rm o}[g_{ab}]$ whose projection to
${\cal M}$ would be concentrated on the possible worldlines of the particle
[see Eq.~(\ref{rQRFcurve})].

Then, diffeomorphism invariance together with imposition of a complex vector
space structure led us to consider general QRFs, $\Psi:{\cal F}({\cal M})\to
{\mathbb C}$: which we have argued can represent not only superpositions of
perspectives but also superpositions of geometries. 

It is important to emphasize, however, that this flexibility does {\it not}
necessarily mean that $\Psi$ must be interpreted as
{\it defining} the superposed geometry
on ${\cal M}$. For it could be that the superposed geometry is already somehow given by some mathematical structure added to $\cal M$, 
and that $\Psi$ merely describes the superposition of ``points of view''
which are
permissible in the given superposed  geometry---which obviously must respect
the fact that these possible perspectives do not belong to a single
orthonormal-frame bundle ${\cal F}_{\rm o}[g_{ab}]$.
In this case too, $\Psi$ does not need to be globally defined, like in the
rQRF case.

In order to avoid confusion and to make it clear when we are dealing with QRFs
which  merely  describe possible superposition of ``points of view'' or  ``perspectives'' on a {\it given}---superposed or 
not---geometry, we refer to such QRFs as {\it perspectival}. So perspectival QRFs do not need to be defined
globally, since they are not  interpreted as defining the  (superposed)
geometry. The ``passive'' character of perspectival QRFs---i.e., complying with,
but not determining nor  
influencing the background---makes them appropriate to describe
superposition of systems whose own gravitational influence (i.e., back-reaction)
is neglected: the so-called ``test'' systems.\footnote{\label{allowdiffusion}{But returning to comment (4), early in Section II: we do allow that a perspectival QRF could obey a continuity-like or Boltzmann-like equation  describing ``diffusion"  in  ${\cal F}({\cal M})$}.}

But agreed: one can envisage, in contrast to this perspectival
conception of QRFs,  a different, more ``basic''  conception of
$\Psi:{\cal F}({\cal M})\to {\mathbb C}$ which {\it does} take $\Psi$ to define and describe  the  superposition of geometries. 
On this conception,
it {\it is} natural to impose that $\pi(supp(\Psi))={\cal M}$---unless 
one were willing to allow pathological scenarios where 
regions of ${\cal M}$  would be left without any geometric structure.
Although the expression ``reference frame'' usually carries the connotation  of being
passive---which is perfectly compatible with the perspectival  conception of QRFs---we shall also use the expression to refer to this  latter, more fundamental
(and possibly dynamical) conception of $\Psi$. So we shall talk of 
{\it basic} QRFs.

But in this paper, we will be  
exclusively interested in  perspectival QRFs, 
because these are  the ones needed to describe probes (like test particles) 
evolving in 
superposition of geometries, which motivated this work in the first place. 
However, the fact that our framework  accommodates, and even 
suggests, this basic conception of QRFs is something which, we believe, should be further 
explored. This would inevitably involve formulating dynamical equations for
$\Psi$, in the same way that GR involves dynamical equations for the metric 
field---and, in fact, Einstein's equation should somehow be recovered
in the context of {\it basic, restricted QRFs}.
But from now on,  we leave such questions aside, and focus on the use of perspectival QRFs
to
describe the scenario that has been a focus in the recent literature on QRFs: the  field due to a macroscopic superposition of a large mass.

\section{A step beyond semiclassical gravity?}
\label{sec:beyond}

 In the previous Sections, we have proposed a  framework 
 where the idea
of QRFs, used in the gravitational context in previous works~\cite{CGBB20,FB22,KHCB22,HKC23}, is formulated in 
a fully
geometric way. Although setting a stage where possible new phenomena can be described---such 
as {\it arbitrary} geometry superpositions---no ``new physics'' has been introduced. 
However, as is often the case, 
the  motivation for introducing new concepts, such as
 QRFs, is the 
prospect of dealing with (at least some) situations which cannot be 
(or are doubtfully) treated with other
available approaches, in the hope that  new
insights can be obtained.

In fact, in 
Ref.~\cite{FB22}, in order to take such a  further step, and propose a new
principle,
the authors consider a paradigmatic situation which is widely believed to
lie beyond the scope of semiclassical gravity (the latter taken to mean a theory where a classical, well-defined spacetime only ``feels''
the average energy-momentum tensor of the quantum matter). 
   Namely: the gravitational field
engendered by a mass in a superposition of two position states---let us vaguely call them 
$\left. \left| L\right. \right> $ and $\left. \left| R\right. 
\right> $, with the latter representing a ``spatial translation''
of the former, from the `Left' to the `Right', by a distance, say, $d$---and the
effect of this field on a test particle.

In this Section, we describe how this scenario is treated by our  framework. We will give a simple description; so this will be a very special case of superposed spacetimes, with each represented by a section of the frame bundle. For we will take the two spacetimes, labelled by $L$ and $R$, to be isometric, i.e.\ the $L$ and $R$ peaks of the mass' spatial wave-function will be idealised as being ``the exact same shape" as each other, and so related by a ``spatial translation.'' (Section \ref{sec:discussion} will return to the general case of superposed spacetimes.) 

\begin{figure*}
\includegraphics[width=0.92\textwidth]{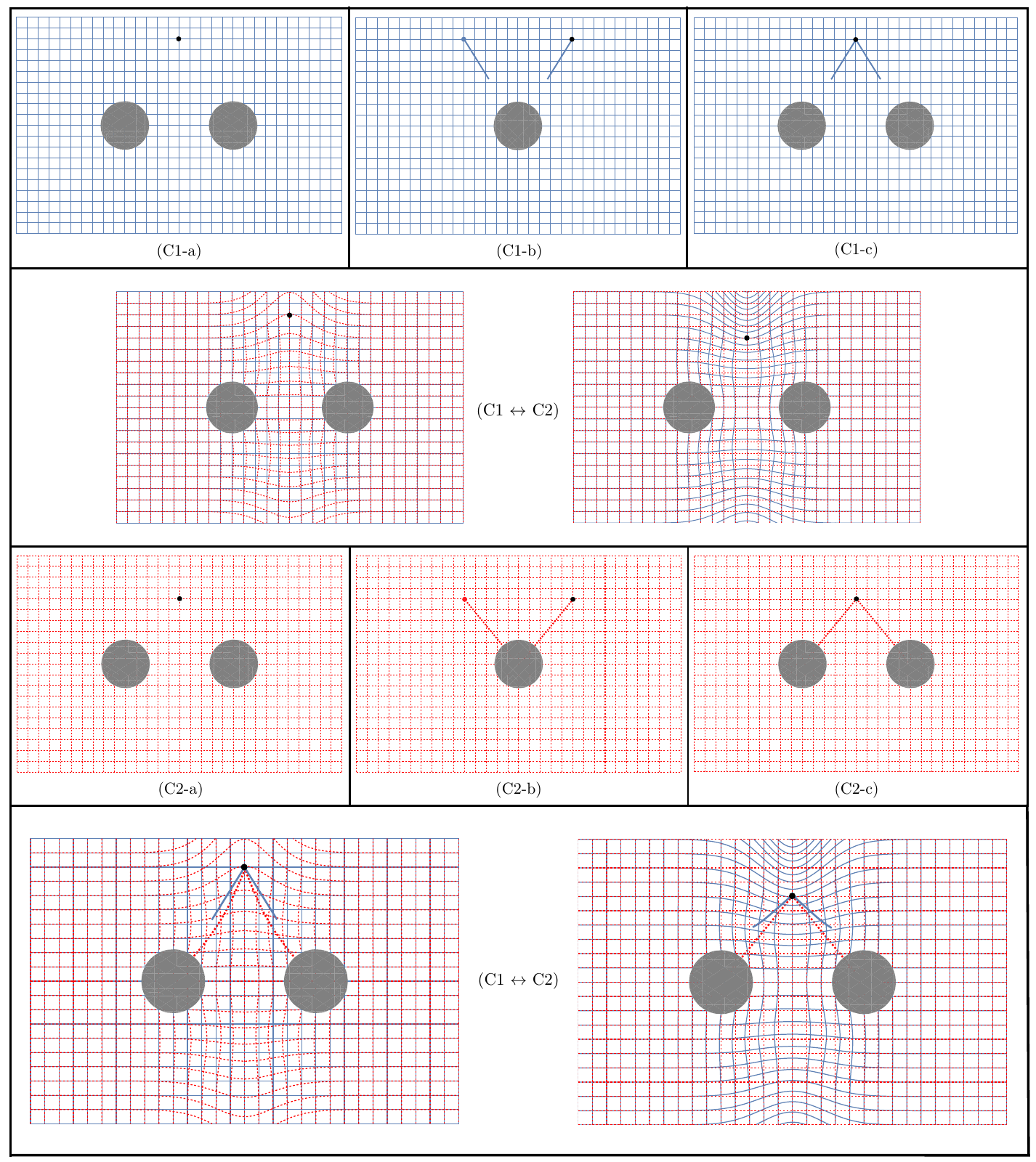}
\caption{We schematically illustrate, here, the
prescription given in Refs.~\cite{FB22,HKC23} for determining the
evolution of a test particle (the black dot) under the influence of
a mass in a macroscopic superposition of position states
(the gray balls). (C1-a) A coordinate system C1 (solid, blue grid) 
is defined on ${\cal M}$, in such a
way that a ``rigid'' {\it coordinate} translation maps one source-mass configuration state into the other (gray balls); (C1-b) After performing 
this coordinate translation, which maps the position state of the test
particle into a superposition of position states (blue and black dots), the 
geodesic equation is solved (for a certain coordinate- or
proper-time interval) for each position state
in the background geometry associated to the now well-defined
source-mass distribution; (C1-c) The inverse coordinate translation is applied to the system (including the solution of the geodesic equation), hence obtaining
a superposition of trajectories (solid, blue curves). (C1$\leftrightarrow$C2--second row) If no well-defined background
metric is assumed for
the superposition of source-mass configuration 
states, there is no way to privilege  coordinate system  C1 in comparison to
another one, such as C2 
(dashed,
red grid). (C2-a,b,c) By applying the same steps above, but now 
using
the coordinate system C2 (with the red dot in C2-b 
playing the same
role as the blue dot in C1-b), 
{\it different} trajectories would be obtained
for the same source-mass superposition state and same test particle
(dotted, red curves).
The relation between these
trajectories is represented in the fourth row. (This figure is
to be seen as a mere qualitative illustration, 
since a more precise depiction should also allow for different choices of coordinate-time ``slices.'')}
\label{fig:Coords}
\end{figure*}

This idealization accords with the recent literature's treatment of the scenario. The difference here, in addition to our representing spacetimes by sections of the frame bundle, is that we will try to state the idealization and its limitations precisely, in terms of reference frames (in our sense, i.e.\ contrasted with coordinate systems). As we warned the reader in the Introduction (concerning unrealistic expectations), we stress here that we do not go beyond
what has been done in the literature using coordinate systems. However, as
we also
said in the Introduction: by  distinguishing physical choices
from arbitrary ones, the geometric approach makes it 
clear that the treatment of this simple
gravitational scenario (coherent superposition of $|L\rangle$ and $|R\rangle$)
using QRFs
lacks predictive power, in the current state of 
knowledge---difficulty which we will also discuss 
in the coordinate-based 
approach in what follows (especially in relation to Fig.~\ref{fig:Coords}), but which, 
we believe, has not been duly
 appreciated in the literature.

If we were to restrict ourselves to the linear regime of 
gravity---which we shall not, to stay faithful to the main motivation
of applying
QRFs to gravitational scenarios---the set-up would be
simple to describe. For there would be an underlying (flat) geometry
where the idea of ``rigid 
translation by a distance $d$'' would make 
clear (i.e.\ unambiguous) sense. However,  the classical spacetimes 
associated
with localized
mass distributions---say
$({\cal M},^{(L)}\!\!g_{ab})$ 
and 
$({\cal M},^{(R)}\!\!g_{ab})$, corresponding to states
$\left. \left| L\right. \right> $ and $\left. \left| R\right. 
\right> $, respectively---are {\it not} translationally 
invariant.
This means that, in general, one cannot change the spatial 
location of a ``test 
system'' of
 $N(>4)$ particles w.r.t.\ the source mass in, say,
 $({\cal M},^{(L)}\!\!g_{ab})$, while satisfying 
 the
 condition that their {\it relative  physical distances} (i.e., the 
 physical distances among the 
 particles) remain unchanged---unless the initial and final positions of the 
 particles are related by an isometry of $({\cal M},^{(L)}\!\!g_{ab})$. 
 Therefore, in general, there is no natural notion of a ``rigid spatial 
 translation'' of an extended system ${\cal S}$ 
 in $({\cal M},^{(L)}\!\!g_{ab})$. From the perspective of the system ${\cal S}$,
 this is equivalent to saying that there is no natural ``rigid spatial translation''
 of the source mass (together with  the geometry it engenders on ${\cal M}$).
 
 The non-existence of a preferred notion of ``rigid spatial translation''
 in  $({\cal M},^{(L)}\!\!g_{ab})$ [and  $({\cal M},^{(R)}\!\!g_{ab})$]  causes difficulties for
 defining a superposition of 
  $({\cal M},^{(L)}\!\!g_{ab})$ and  $({\cal M},^{(R)}\!\!g_{ab})$.
For  if such a notion existed, it could be used to 
 naturally
 identify points in $({\cal M},^{(L)}\!\!g_{ab})$ with points in $({\cal M},^{(R)}\!\!g_{ab})$
in a nontrivial way (i.e., not through the  
 isometric identification which maps source to source). Namely, in such a way that a given point of the manifold with the superposition would correspond to points in both
   $({\cal M},^{(L)}\!\!g_{ab})$ and  $({\cal M},^{(R)}\!\!g_{ab})$ with 
   possibly  different local properties.
In fact, the protocol for such identification would be: (i)~since $({\cal M},^{(L)}\!\!g_{ab})$ and
 $({\cal M},^{(R)}\!\!g_{ab})$ are isometric, the hypothetical 
 {\it displaced-to-the-left} ${\cal S}$ in $({\cal M},^{(L)}\!\!g_{ab})$ could be isometrically mapped to points of $({\cal M},^{(R)}\!\!g_{ab})$; (ii)~then,
 the identification is made between the original (undisplaced) ${\cal S}$  in 
 $({\cal M},^{(L)}\!\!g_{ab})$ and the image, in $({\cal M},^{(R)}\!\!g_{ab})$,
 of the displaced-to-the-left ${\cal S}$; (iii)~finally, by considering arbitrary
 ${\cal S}$---for instance, filling in each entire spatial section with constituents
 of the system ${\cal S}$---one 
 would get the
 desired
 identification between  $({\cal M},^{(L)}\!\!g_{ab})$ and  $({\cal M},^{(R)}\!\!g_{ab})$. With such an identification, ${\cal S}$ would be seen to be ``the same''
  in both spacetimes---hence, each constituent of ${\cal S}$ would be subject to two
 different local geometries: a {\it superposition}.

This is an important
point which must be treated carefully.
In Refs.~\cite{FB22,HKC23}, the authors try to address this point by fixing a spatial
coordinate
system in the superposed configuration
in such a way that a 
{\it coordinate} translation maps
the mass distribution
associated to $\left. \left| R\right. 
\right> $  into the mass distribution associated to 
$\left. \left| L\right. 
\right> $---hence, making the background geometry well defined after
this ``matching'' translation is performed. This is just the same
identification protocol mentioned above for an everywhere-defined 
``system'' ${\cal S}$; but now with a coordinate system playing the role of 
${\cal S}$, and with the constraint of preserving relative 
{\it physical} distances---which, in general, cannot be 
satisfied---weakened to requiring just preservation of relative ``coordinate distances'': which, by construction, is trivially satisfied for {\it any} coordinate system 
globally-defined first
 on $({\cal M},^{(L)}\!\!g_{ab})$. As a result, there are infinitely many {\it different} coordinate systems defined on the superposed configuration
which can be equally well
adopted for carrying out  
this
protocol (if no further conditions are imposed). And
different choices 
would lead to
different mappings, eventually leading to different {\it physical} conclusions---which would be unacceptable: see Fig.~\ref{fig:Coords}.

A possible 
way out of this conundrum would be to select  the  coordinate system 
according to
a clear physical prescription. However, this cannot be done
in the superposed mass configuration 
without  risking circular reasoning.
For we do not know, beforehand, the ``physics'' of the background 
associated to such a superposition; and
there is no reason to privilege one or the other  ``branch'' of the geometry. 

One might try, for example, fixing a physical coordinate system
{\it before} the superposition is prepared and then consider its ``evolution.''
 In Ref.~\cite{HKC23}, 
this is attempted by saying that a Euclidean 
coordinate system is ``fixed'' in the laboratory 
before the superposition is prepared---perhaps even 
before the mass is brought
in for the experiment, 
since the spacetime is assumed to be flat then. However, in a spacetime which 
is non-stationary---which it must be, otherwise it would continue to be 
flat---there is no natural identification
which allows one to say which are ``the same points in space'' before and after the mass is brought in (and then superposed): in a non-stationary geometry, 
the idea of ``holding things in place'' so as to set the spatial coordinates
is meaningless. 
That is:  
in the non-linear regime, the spatial 
coordinate system defined in such a way  that the
nearby objects in the ``laboratory'' are ``spatially fixed'' 
would strongly depend on
the details of the stresses
and strains in the physical objects constituting the 
``laboratory''. (This is just the physical counterpart of
our discussion above that there is no natural notion of spatial translation in a
background geometry which is not translationally invariant.) 
And, again, we risk incurring circular reasoning since we do not
know the physics of the  geometry's {\it evolution} 
 from an initially well-defined spatial geometry up to the 
 geometry associated with   the mass in superposition.

This point is, indeed, intricate. And agreed: here we shall not be able to do 
better than Refs.~\cite{FB22,HKC23}, simply by recasting the problem in geometric 
terms---except, perhaps,  in recognizing its subtleties and 
pointing to its possible {\it dynamical} nature: as  hinted at above,
with our discussion of the evolution of physically-constructed coordinate systems,
and further discussed in what follows.

In our geometric set-up, 
we may take ${\cal M}$ to be the arena where the superposition is to
be defined---hence points of the manifold are naturally identified in both
``branches'' of the geometry---and the challenge is to ascribe the metric fields 
$^{(L)}\!g_{ab}$ and $^{(R)}\!g_{ab}$ to ${\cal M}$ in such a way that
each represents the spacetime of each position configuration of the source mass.
For a start,
we
want  $({\cal M},^{(L)}\!\!g_{ab})$ and 
$({\cal M},^{(R)}\!\!g_{ab})$ to
be isometric: $^{(R)}\!g_{ab} =
\phi_d^\ast\ \!\!^{(L)}\!g_{ab}$, with $\phi_d$ being a diffeomorphism
on ${\cal M}$
which should satisfy some conditions in addition to mapping the
mass distribution of $\left. \left| L\right. \right>$ into
that of $\left. \left| R\right. \right>$.
Since the ``relative'' separation of the positions associated to the 
states $\left. \left| L\right. \right> $ and $\left. \left| R\right. 
\right> $ is supposed to be fixed, $\phi_d$ should
preserve the static worldlines: $\phi_d^\ast 
\xi^a_{(L)}\propto \xi^a_{(L)}$, where 
$\xi^a_{(L)}$ is the Killing field representing the static symmetry of 
$({\cal M},^{(L)}\!\!g_{ab})$.
Moreover, since $({\cal M},^{(L)}\!\!g_{ab})$ is asymptotically flat at
spatial infinity, we can demand  $\phi_d$ to ``tend'' (in a sense which must be
made precise)
to rigid spatial translations as we consider points arbitrarily far
away from
the mass distributions.\footnote{In a more realistic situation, where the superposition is obtained from
a previously well-defined mass distribution, one should impose that $\phi_d$ is trivial outside the causal future of the region where the superposition was prepared. But here we consider the idealized case where the superposition has always existed.}
We  may also impose, for symmetry purposes, 
that the dependence of $\phi_d$ on $d$ (as a one-parameter family of
diffeomorphisms) is 
such that $\phi_d^{-1} = \phi_{-d}$.

All these conditions, however, 
do not fix a unique $\phi_d$; in fact,
they still leave an infinite number of possibilities---which is the 
geometric counterpart of the infinitely many coordinate systems
mentioned above.
And it is relevant to note that, 
even though different possibilities of $\phi_d$ lead to spacetimes
$({\cal M}, \phi^{\ast}_d\ \!\!^{(L)}\!g_{ab})$ which are physically
indistinguishable among themselves,
superpositions of $^{(L)}\!g_{ab}$
and $^{(R)}\!g_{ab} = \phi^{\ast}_d\ \!\!^{(L)}\!g_{ab}$  {\it are sensitive} to
$\phi_d$, in accordance with our previous discussion that
different choices of coordinate systems (for which a coordinate translation maps
$|R\rangle$ to $|L\rangle$), which determine different choices of $\phi_d$, lead
to different physical conclusions.
In other words, 
determining {\it the} diffeomorphism $\phi_d$ is a
{\it physical} question.

Here, we pragmatically assume that  one 
such
diffeomorphism has been privileged. Based on our previous discussion
about the evolution of physically-constructed 
coordinate systems, it is natural to
conjecture that such a diffeomorphism should be selected
by details of the background
{\it dynamics} which evolved  the total system so as to prepare
the superposition. So, whatever perspectival QRFs
we use to describe quantum probes on this superposition, they should all
be consistent
with the basic QRF  which, we presume, describes the superposed geometry.\footnote{We say that a perspectival QRF $\Psi_P$ is {\it consistent} with a basic QRF $\Psi_B$ if for each $(p;\{{\bf x}_\mu^a\})\in supp(\Psi_P)$, there is at least one $\Lambda \in O(3,1)$ such that $(p;\{(\Lambda{\bf x})_\mu^a\})\in supp(\Psi_B)$.} This dynamics could perhaps be characterized by 
one-parameter families of diffeomorphisms,
$^{(L)}\!\phi_t$ and $^{(R)}\!\phi_t$, each describing the evolution of
each ``branch,'' with $\phi_d :=\, ^{(R)}\!\phi_t 
\circ \, ^{(L)}\!\phi_t^{-1}$ for $t$ sufficiently large 
to relate the asymptotic static configurations.   But at this point, we have no detailed suggestion about this dynamics, or the diffeomorphisms
$^{(L)}\!\phi_t$ and $^{(R)}\!\phi_t$.

Given one preferred $\phi_d$,  the spirit of our approach 
to the QRF idea
is
that the {\it geometric effects} of
the superposition
$\alpha|L\rangle+
\beta|R\rangle$---with each of $|L\rangle$ and
$|R\rangle$ associated to 
$({\cal M},^{(L)}\!\!g_{ab})$ and $({\cal M},^{(R)}\!\!g_{ab})$, respectively, with $^{(R)}\!g_{ab} = 
\phi^{\ast}_d\ \!\!^{(L)}\!g_{ab}$---on a ``test system'' (i.e., one which can 
be neglected as a source of changes to the  geometry)
can be described  by adopting the perspectival QRF $\Psi = \alpha\Psi_L+\beta\Psi_R$, where $\Psi_L$ and $\Psi_R$ are {\it arbitrary}
rQRFs defined on $({\cal M},^{(L)}\!\!g_{ab})$ and $({\cal M},^{(R)}\!\!g_{ab})$, respectively.

Note that, on their own, $\Psi_L$ and $\Psi_R$ represent (possibly
different) ``points of view'' (i.e.\ frames) 
on isometric spacetimes.
However, $\Psi$
itself is not a rQRF on any classical spacetime.  And as such, it
represents, in principle, a completely different
situation, beyond the scope of our current (classical-
and semiclassical-gravity) understanding.

Here is where a new {\it invariance} principle is introduced. 
According to our reading of the literature, it is the implementation of the so-called ``quantum covariance'' proposed in the coordinate-based approach---see, e.g.,
Sec.~IIA of Ref.~\cite{HKC23}; but we now rephrase in terms of the geometric language presented here.
Although, as pointed out above, $\Psi = \alpha\Psi_L+\beta\Psi_R$ describes a situation beyond classical or semiclassical 
gravity, it is {\it imposed} that it should lead to the same observables as  (i.e., be ``equivalent'' to) the {\it {\it restricted}  QRF} 
$\widetilde{\Psi}$ 
obtained using the pullback of
$\Psi_R$ to $({\cal M},^{(L)}\!\!g_{ab})$:
\begin{eqnarray}
\widetilde{\Psi}(p;\{{\bf x}\}) := 
\alpha \Psi_L(p;\{{\bf x}\}) + \beta 
\Psi_R(\phi_d(p);\{{\phi_{d}^{\ast}{\bf x}}\}) : 
\label{PsitildeA}
\end{eqnarray}
the values of  observable quantities should be {\it invariant}
under $\Psi\mapsto \widetilde{\Psi}$.  With $\widetilde{\Psi}$
representing a mere  ``point of view'' on a well-defined background 
geometry---namely, $({\cal M},^{(L)}\!\!g_{ab})$---this brings the description of the evolution
of the test system within the jurisdiction of classical or semiclassical gravity.\\

This is the analogue, in the gravitational scenario, of the ``transfer'' of the superposition
of the state of the system to a superposition of ``points of view'' which we illustrated in Subsec.~\ref{subsec:flat} in the 
flat-spacetime context. Here, the superposition of geometries (which is the ``physical system''), described by $\Psi$, is mapped
to a superposition of ``points of view'' in a non-superposed geometry, described by $\widetilde{\Psi}$. Since $\Psi_L$ and $\Psi_R$ are arbitrary, they can, in particular, {\it each}  correspond to  a ``point of view'' of a localized test particle; cf.\ again row 1 i.e.\ C1 of Fig.~\ref{fig:Coords}.

For instance, in the case of a free (point-like)
test particle with a given initial condition---say, passing through
 the manifold point $p_0$ 
with tangent vector $v_0^a\propto \xi^a_{(L)}$, at its
{\it proper time} $\tau_0$---as
considered in Ref.~\cite{HKC23}, 
we may consider $\Psi$ to be the ``point of view'' of the particle
itself, in the sense of Eq.~(\ref{rQRFcurve}) (but without our needing here
to adopt any coordinate system); i.e., both $\Psi_L$ and $\Psi_R$
have supports whose projection to ${\cal M}$ are worldlines (to be determined) ``starting'' at $p_0$ 
with tangent vector $v_0^a$. Hence, $\widetilde{\Psi}$ given by 
Eq.~(\ref{PsitildeA}) has support whose projection to ${\cal M}$ 
are worldlines (still to be determined) 
``starting'' at $p_0$ and $\phi_d^{-1}(p_0)$ with tangent vectors
$v_0^a$ and $\phi_{d\ast}v_0^a$, respectively.
But now 
$\widetilde{\Psi}$ 
is a {\it restricted} QRF on 
$({\cal M},^{(L)}\!\!g_{ab})$; so, the condition that the particle is free means that its possible worldlines 
are the 
geodesics of $({\cal M},^{(L)}\!\!g_{ab})$
determined by the initial conditions
$(p_0,v_0^a)$ and $(\phi_d^{-1}(p_0),\phi_{d\ast}v_0^a)$---call them $\gamma_1$
and $\gamma_2$, respectively. 
Then, 
going back to the initial QRF $\Psi$, we have, in this case,
$\pi\left(supp(\Psi)\right) = \pi\left(supp(\Psi_L)\right)\cup 
\pi\left(supp(\Psi_R)\right)=
\gamma_1\cup \phi_d[\gamma_2]$, where $\pi:{\cal F}({\cal M})\to {\cal M}$ is as before  the
canonical projection defined on the bundle of frames. 

This is just the expected result obtained in the coordinate-based formulation of QRFs: the particle evolves to a superposition of geodesics,  
referring to each $({\cal M},^{(L)}\!\!g_{ab})$ and 
$({\cal M},^{(R)}\!\!g_{ab})$---bearing explicit dependence on the
diffeomorphism $\phi_d$.
A figure representing the procedure described above (projected to the
static
spatial sections) would 
be quite similar to Fig.~\ref{fig:Coords} with {\it one} given
preferred choice of coordinate system.

\medskip

Note that 
Eq.~(\ref{PsionF})---which
is just {\it classical} diffeomorphism invariance 
expressed in the context of QRFs---ensures that the principle
expressed by Eq.~(\ref{PsitildeA}) is actually independent of the
``representative'' $({\cal M},^{(L)}\!\!g_{ab})$ with which we start
our construction. We could instead have started with $({\cal M},^{(R)}\!\!g_{ab})$
or any other isometric spacetime. A more symmetric---but 
equivalent---description, for
instance, would be obtained by defining $\widetilde{\Psi}$ using the pullback
of both $\Psi_L$ and $\Psi_R$ through the hypothetical 
$^{(L)}\!\phi_t$ 
and $^{(R)}\!\phi_t$ mentioned above, 
respectively, for $t$ sufficiently large to
relate the asymptotic static configurations.
 
 \medskip
 
The same rationale can 
be applied 
to analyze a static (point-like) ``clock''---as also considered in
Ref.~\cite{HKC23}---whose proper time 
will then evolve as a superposition of two proper times.

 These examples about worldlines and proper times illustrate that---at least as far as geometrical observables are concerned (of which geodesics and proper times are examples)---any conclusion one might reach using the original coordinate-based
formulation of QRFs applied to the simple gravitational scenario of superposition of well-defined spacetimes, as in Refs.~\cite{CGBB20,FB22,KHCB22,HKC23},
can be obtained using the geometric formulation
we have presented here. At least, this is true, provided that the subtlety 
which we highlighted above, about the non-uniqueness of
the so-called ``spatial translation'' relating $|L\rangle$ and
$|R\rangle$, is
properly addressed in {\it both} formulations.

\section{Discussion}
\label{sec:discussion}

We have presented a  geometric formulation of the idea of a 
QRF, which had already been applied,
in a coordinate-based way, to describe the effects of
gravitational fields engendered by masses in coherent
superpositions
of position states~\cite{HKC23}.
Our approach has been to take a QRF as
a mapping $\Psi:{\cal F}({\cal M})\to {\mathbb C}$ (seen as  ``wave
functions'' on the fibres of ${\cal F}({\cal M})$); and thus to rely heavily on the conceptual distinction between
``points of view'' (on which values of observable quantities {\it can} depend) and 
coordinate systems (whose different choices 
can make no difference to
observed values).

In a nutshell, the  idea of QRFs
applied to the gravitational scenario is
simple. A ``test''  system (i.e., one whose own gravitational effects can be 
completely neglected) would supposedly evolve as a coherent superposition of ``histories,'' one for each well-defined position state of the
source mass---at least as long as it does not disturb the very
state of the
source mass.
Although some may consider that this ``trivializes''  the effect of a superposition of geometries,
the idea of QRFs  formalizes calculations without the need for a well-defined 
background metric---once a preferred coordinate system
or, equivalently, a preferred diffeomorphism relating both
source-mass configurations is selected.

We emphasize that our purpose here has not been to assess the merits of the general idea of QRFs, but merely to recast it in a geometric
language. By doing so,
it is our view that some conceptual aspects of
its 
{\it implementation}, particularly 
when going beyond the
linear-gravity regime, 
get highlighted---such as 
the {\it necessity} of a  preferred diffeomorphism 
relating the source-mass configurations and
its possible relation with the (unknown, quantum)
dynamics which prepares the superposition. In fact, although we have mentioned, in the Introduction, that other (somewhat related) 
coordinate-independent approaches have come to our attention at a late stage of this work---specifically Refs.~\cite{HGHLM21,JG24}---to 
the best of our knowledge none of them (being 
top-down approaches)
has been applied to the
concrete scenario treated in
Refs.~\cite{CGBB20,FB22,KHCB22,HKC23} and here (Sec.~\ref{sec:beyond}). As a consequence, none of them
have been used to make the point raised here in Sec.~\ref{sec:beyond}, namely:
that in the current state of knowledge, i.e.\ without further (as yet unknown) input from physics, the idea of a QRF lacks predictivity about gravitational scenarios, even very simple ones such as the superposed mass we have considered.

%\\
%{\jtod On Zoom on Friday5July, you spoke as if you might make the wording of the above paragraph stronger, i.e. a bit more critical of the literature. I leave it to you! Happy with whatever you decide.}

\bigskip

Having just mentioned  other coordinate-independent approaches, we shall now  briefly compare our formulation with  the very recent  proposal  presented in Ref.~\cite{JG24}. Like this paper, it models a classical reference frame as a tetrad field. Then it models a quantum reference frame as, roughly speaking, a quantum system (with Hilbert space ${\cal H}_{\cal R}$)
equipped with a covariant mapping of (measurable subsets of)
tetrads into effects on ${\cal H}_{\cal R}$.\footnote{An effect is a positive operator whose spectrum is a subset of $[0,1] \subset {\mathbb R}$. Thus it is generalization of a projection operator; and in operational quantum physics (cf.\ e.g.\ Ref.~\cite{BGL95}), the traditional conception of an observable as a projection valued measure (PVM) (e.g.\ on $\mathbb R$: mapping Borel subsets of $\mathbb R$ to projection operators) is generalized to a positive operator valued measure (POVM).} Unfortunately, this paper came to our attention too late for us to undertake a proper comparison with it, close cousin of this paper though it is. But let us give a short discussion.

 Ref.~\cite{JG24} belongs to the {\it operational} approach to quantum reference frames (cf.\ e.g.\ Refs.~\cite{LMB18,CGL23}). For 
our purposes, the main general ideas of the operational approach are that: 
(i)~a quantum reference frame is taken as a quantum system (hence with its own Hilbert space), together with a quantity on it (called the {\it frame observable}: taken as an appropriately covariant POVM); (ii)~quantum uncertainty as to the state of the frame is modelled by the frame’s state determining a (orthodox Born-rule) probability for that quantity’s values; and (iii) the description of the object-system is of course sensitive to the frame’s quantum state and the frame observable. 
[Here, (i)-(iii) correspond to Ref.~\cite{JG24}’s Eqs.~(1), (6), and (11)]. Ref.~\cite{JG24} then adds to these general ideas the proposal we hinted at above. Namely: a quantum reference frame---more precisely, its frame observable---is given by a POVM mapping appropriate (Borel) subsets of the frame bundle to positive operators on the Hilbert space of the frame. (Cf.\ Ref.~\cite{JG24} Section 5.1 and 5.2, especially Eq.~(60) which generalizes its Definition 4.1.)

This invocation of tetrads makes the bridge to our framework. In particular, Ref.~\cite{JG24}’s frame observables [and especially (iii) above: i.e.\ the description of the object-system being sensitive to the frame’s quantum state and observable: cf.\ the relativization map on quantities, 
Eq.~(13)] thus becomes an abstract and general cousin of our framework’s coupling of tetrads to the object-system fields, as in our discussion of our Eq.~(\ref{EvalueQRF}).
It is also noteworthy that both frameworks use the frame bundle as a way to treat superpositions of geometries (cf.~the start of our Sec.~\ref{sec:genQRF}). 
But a more detailed comparison of the two frameworks must be postponed to future work.

\bigskip

We also make no claim that the geometric formulation presented here
is ``minimalist'' in any sense. In fact, since ``points of view''
may affect values of observable quantities but (we presume) 
{\it not} the  evolution of the system itself, one
might try to devise a simpler or ``more economic''
geometric formulation by, e.g., doing away 
with (or ``tracing out'') 
different points of view and focusing solely on
spacetime geometries---i.e., basing the construction on 
the ``space'' of possible
 geometries instead of on the bundle of 
frames. 

This last comment prompts a comparison between this paper’s 
fibre-bundle framework and that of Ref.~\cite{KHACGBB24}  (cf.\
also Refs.~\cite{GB23, GB23a}, and Ref.~\cite{G24a} Sections 3, 7 and Appendix A).\footnote{The two frameworks were developed independently; (this paper’s framework,  mostly by D.V.)
We thank the authors of these papers, with whom  later discussions  have helped us understand the relation between the frameworks.}

It will be clearest to begin with an obvious contrast, about the dimensions of the fibre bundles. This will help us to spell out how the two frameworks differ in their treatments of how to identify spacetime points across two different spacetimes.\footnote{\label{knit5and6} Recall Section~\ref{sec:beyond}’s discussion of how to make rigorous sense of ``which point is which” when comparing two spacetimes. There, we first saw that if one spacetime admitted a notion of rigid spatial translation by distance $d$, this notion naturally defined a protocol for identifying points. But we then stressed that such a notion is {\it not} in general available, and thus we discussed what features a diffeomorphism ${\phi}_d$ might be hoped or required to have, for it to effect such an identification.}

Our fibre bundle ${\cal F}({\cal M})$ is finite-dimensional, with the spacetime manifold $\cal M$ as its 4-dimensional base-space and with the bases at each spacetime point $p$ forming the 16-dimensional fibre above $p$. 
(And similarly, for 
Section~\ref{sec:QRFrestricted}’s fibre bundle ${\cal F}_{\rm o}[g_{ab}]$ of orthonormal frames of $({\cal M}, g_{ab})$, whose
fibres are 6 dimensional.) But in the fibre bundle adopted by Ref.~\cite{KHACGBB24}, both (i)~the base-space, and (ii)~each fibre, are infinite-dimensional. Namely, for vacuum general relativity on the manifold $\cal M$, they are: (i)~the set of Lorentzian geometries, where each geometry is determined by an isometry-class of Lorentzian manifolds\footnote{{Strictly speaking, this brief statement needs to be qualified so as to register the role of four scalar fields that one has to choose; cf.\ the discussion below.}} 
$({\cal M},g_{ab})$, and (ii)~such an isometry-class, i.e.\ an orbit of the diffeomorphism group Diff(${\cal M}$) of $\cal M$. (Cf.\ the discussion at the start of Sec.~\ref{sec:genQRF}.)

This difference implies immediately that: (a)~for us, a spacetime with a specific metric (and equipped with a tetrad field) is given by a section of ${\cal F}({\cal M})$, so that two such spacetimes in superposition are represented by a (basic) QRF $\Psi$ with support on two such sections; 
(more precisely:   on two such sections that are not related by an application of an element of $O(3,1)$ at each spacetime point $p$ of the manifold $\cal M$). On the other hand: (b)~for Ref.~\cite{KHACGBB24}, a spacetime with a specific metric (and {\it not} equipped with a tetrad field) is a point, i.e.\ element of the bundle, so that two spacetimes in superposition are to be represented by two such points (each with a complex amplitude).

This difference also means that: (a)~for us, identifying a point across two spacetimes in superposition is simply a matter of ``looking up or down along the  fibre, from one section to the other”, since the fibre is labelled by the point $p$. So for us, the self-same point has different metrical (and material) properties and relations to other points (as encoded by the metric and matter-field tensors) on the different sections. 

On the other hand,  (b):~for  Ref.~\cite{KHACGBB24}, identifying a point across two spacetimes in superposition is a matter of ``looking inside” the points of the fibre bundle (each of which is an entire spacetime, endowed with a configuration of
four  coordinate scalar fields---the so-called $\chi$-fields), then focusing on the {\it spacetime points} therein, and then defining  
what it is for
two points $p$ and $q$ in non-isometric 
spacetimes, or in isometric spacetimes with different 
$\chi$-fields configurations,
to be identified with each other, so as to move from one fibre to another.

In Refs.~\cite{KHACGBB24, GB23, GB23a}, two points that are thus identified are said to be {\it threaded}. In Ref.~\cite{GB23a}, and Ref.~\cite{G24a} (Sections 3, 7 and Appendix A), the relations between threading and the topic of connections on fibre bundles is discussed. In Ref.~\cite{KHACGBB24}, this threading is done by: (i)~making a choice of four scalar fields (the $\chi$-fields)
that take, in a single spacetime, suitably non-repeating values, so that any two points of $\cal M$ (or working locally: of a region ${\cal O} \subseteq {\cal M}$) have distinct quadruples of values; and then (ii)~saying that two points $p$ and $q$ in ``distinct'' spacetimes are to be threaded (i.e.\ are to be identified: ``are physically the same”) iff their quadruples of values for the four scalar fields match. (Obviously, (i) and (ii) here are a formal analogue or model of Section \ref{sec:beyond}’s considerations about what features a diffeomorphism ${\phi}_d$ might be hoped or required to have, for it to effect such an identification. Cf.\ footnote~\ref{knit5and6}. Hence also our scare-quotes around the word `distinct': for the spacetimes do not need to be non-isometric---they are not, in Section \ref{sec:beyond}’s case of the mass in a macroscopic superposition of position states.)

 These different treatments of how to non-isometrically identify points across spacetimes 
  prompt a brief philosophical comment.  Despite the difference, we think {\it both} treatments are compatible with the moral drawn from the hole argument at the start of Sec.~\ref{sec:genQRF}, viz.\ that points only acquire significance through the physical quantities evaluated at them. 
 
 As to (a): this paper’s treatment in effect makes points’ identity {\it fiducial}. For nothing turns on whether the point at the base of (and so labelling) a fibre is $p$ or some other point, say $q$---it is just that in order to define the bundle, a choice must be made.\footnote{This is reminiscent of the philosophical doctrine called `haecceitism’, whose tenability as a response to the hole argument was first formulated by Ref.~\cite{B89} (p.\ 21, Sec.\ 5).}

 And (b): we of course agree that Ref.~\cite{KHACGBB24}’s framework of identification-by-matching-field-values is also compatible with this moral from the hole argument. This compatibility is also reinforced by:\\
\indent (1) this framework’s admission that the scalar fields in (i) above involve a choice, i.e.\ that other quartets of suitably non-repeating scalar fields are equally legitimate; and \\
\indent (2) this framework’s having a motivation (cf.\ Ref.~\cite{GB23}) given by a philosophical theory about identity, called {\it counterpart theory}.

So much by way summarising the contrast between the two frameworks about dimensions, and their ensuing differences about how to identify points across spacetimes. Finally, we note an obvious but important way in which the frameworks are concordant. The underlying point is that---as all must agree---the set of spacetimes is infinite-dimensional. This is of course explicit in the fibre bundle adopted in the framework of Refs.~\cite{KHACGBB24,GB23, GB23a}. But of course, it is also true, though implicit, in our framework. 
In fact, if we were to build the classical ``space of models''
(as the bundle of 
Ref.~\cite{KHACGBB24} is called) based on our framework (but without any reference to the complex amplitudes $\Psi$ yet), we would have to work
with the space ${\text{\it Sec}}[{\cal F}({\cal M})]$ of local sections of ${\cal F}({\cal M})$---which is of course infinite-dimensional as well.

Thus we stress that in this way, the dimension contrast which we have expounded does not involve any conflict between the two frameworks. Agreed: in this way it could seem a bit odd that we have compared our finite-dimensional bundle ${\cal F}({\cal M})$ with the infinite-dimensional bundle of 
Ref.~\cite{KHACGBB24}. But our motivation for making this comparison is clear: in each framework, these are the bundles which function as the stage on which to define QRFs---with the infinite-dimensional space ${\text{\it Sec}}[{\cal F}({\cal M})]$ playing no  role here.

\medskip

But we must leave further comparisons of our framework with other frameworks {\j especially those of Ref.~\cite{JG24} and Ref.~\cite{KHACGBB24}} to future work.
We end this Section with a summary, using a Figure, of how this paper's framework arises from a handful of ideas; and this summary will lead to a brief final list of open questions.

\begin{figure*}
\includegraphics[width=\textwidth]{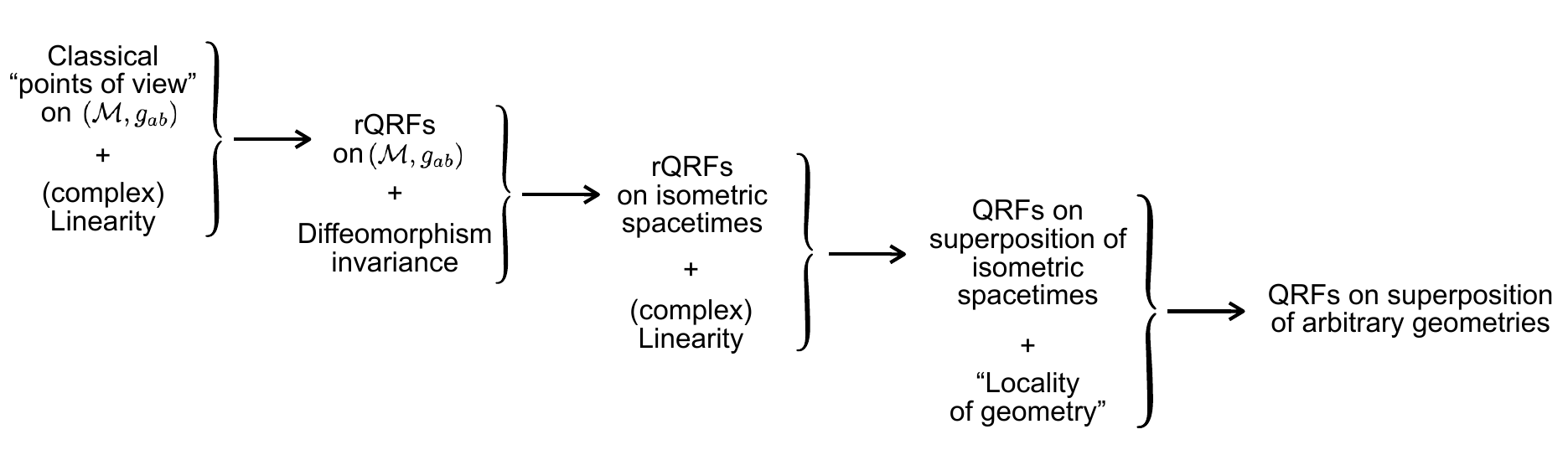}
\caption{Schematic rationale for our geometric formulation
of QRFs (presented in Secs.~\ref{sec:QRFrestricted} and
\ref{sec:genQRF}).
It starts  (in Sec.~\ref{sec:QRFrestricted}) 
as a way to superpose (using assignments of complex amplitudes) 
different ``points of view'' on a given spacetime (rQRF). Then, 
imposing diffeomorphism invariance (in Sec.~\ref{sec:genQRF}) 
naturally leads to maps between
rQRFs on isometric
spacetimes. Requiring (also in Sec.~\ref{sec:genQRF}) 
superposition to again make sense
(via assignments of complex amplitudes), we are ``forced'' to consider 
superposition of ``equivalent'' (global) 
geometries (i.e., isometric spacetimes). However, given an open neighborhood ${\cal O}\subsetneq {\cal M}$, it is {\it not} possible, in general, to distinguish isometric from non-isometric {\it global} geometries by their restrictions to
${\cal O}$. Therefore, if we do
{\it not} want 
to impose constraints on $\left.\Psi\right|_{\cal O}$ which would depend on 
global structures (a sort of ``locality'' principle), then we are 
naturally 
led (still in 
Sec.~\ref{sec:genQRF}) to consider  superpositions of {\it arbitrary} geometries.}
\label{fig:QRF}
\end{figure*}

We  believe that this paper's framework follows naturally from combining the core idea of QRFs with some well-established general ideas, namely 
linearity, diffeomorphism invariance, and what we will call `locality of geometry'. 
The latter means that there is no way to determine 
whether two spacetimes (defined on the same manifold ${\cal M}$)
are isometric by only ``looking at'' an open region ${\cal O}\subsetneq 
{\cal M}$. Hence, if a QRF $\Psi$ is to be locally constructed, 
there is no reason to consider only superpositions of globally isometric spacetimes;
superposition of arbitrary geometries must be allowed.
In Fig.~\ref{fig:QRF}, we depict, in a schematic way, the line of reasoning which motivated our framework.

\medskip

As with any mathematical
framework in physics, the
merit of our geometric formulation of QRFs
must be assessed
by how useful it can be for portraying known scenarios and/or
enabling the description and understanding of new ones. We note in particular that this paper has {\it not} explored the generality enabled by taking a QRF as a map $\Psi:{\cal F}({\cal M})\to {\mathbb C}$:
which allows
{\it arbitrary} superpositions of spacetime geometries (on 
${\cal M}$). 
For in Sec.~\ref{sec:beyond}, 
our reading of the (new) invariance principle which underpins
 the use of QRFs in the gravitational scenario analyzed in Refs.~\cite{FB22,HKC23}---viz.\ the 
 equivalence of
$\Psi = \alpha \Psi_L +\beta \Psi_R$ and
$\widetilde{\Psi}$ given by Eq.~(\ref{PsitildeA}), 
for $supp(\Psi_L)\subseteq {\cal F}_{\rm o}[^{(L)}\!g_{ab}]$
and $supp(\Psi_R)\subseteq {\cal F}_{\rm o}[^{(R)}\!g_{ab}]$,
with $^{(R)}\!g_{ab} = \phi^{\ast}_d\ 
\!\!^{(L)}\!g_{ab}$---was applied only to the very special
cases where $\Psi$ describes the superposition of two (but easily
generalizable to a finite number of) {\it isometric}
spacetimes.

Thus we arrive at some natural questions for future analyses. Can this invariance principle be extended to a more
generic $\Psi$? More importantly, can such  generality be put 
to some
use beyond mere formal description? At a higher degree of
speculation: can this framework be used to say anything about 
the 
{\it evolution} $\phi_t$ hypothesised in 
Sec.~\ref{sec:beyond}? 

There are also interesting questions about the 
{\it meaning}
of the QRF assignments 
$\Psi:{\cal F}({\cal M})\to {\mathbb C}$, and even
of the rQRFs $\Psi:{\cal F}_{\rm o}[g_{ab}]\to {\mathbb C}$. Here, we recall the Section \ref{sec:kd}'s distinction between perspectival and basic conceptions of a QRF.
For an rQRF $\Psi:{\cal F}_{\rm o}[g_{ab}]\to {\mathbb C}$, where the spacetime metric
is well defined, the classical principle of {\it general
covariance} demands that the {\it physics} of the systems being 
described with respect to a particular 
$\Psi:{\cal F}_{\rm o}[g_{ab}]\to {\mathbb C}$ cannot depend on
the choice of $\Psi$. This demand is not to be confused with values of observables:
which {\it do} depend on $\Psi$, but in such a
way that the physical system being ``observed''
has an absolute, frame-independent underlying evolution. For this 
reason, the question of whether the rQRF $\Psi$ represents a physical 
apparatus or a mere fictitious idealisation was irrelevant. 

However,
in the general case, different choices of $\Psi:{\cal F}({\cal M})\to 
{\mathbb C}$ {\it can} represent different physical situations; 
and so it is more reasonable that $\Psi$ itself should satisfy
further physical constraints, such as continuity-like and/or 
Boltzmann-like transport equations on ${\cal F}({\cal M})$.

Again at a higher degree of speculation: could
time-interval and distance uncertainties introduced
by such a $\Psi$
be (at least partially) 
responsible for the uncertainty relations which are 
usually
obtained from the non-commutativity of observables?
And more generally, can the formalism of describing arbitrary superpositions of geometries by $\Psi:{\cal F}({\cal M})\to 
{\mathbb C}$  have applications other than for the QRF idea?

The answers to these questions are unclear. But
we hope that
our fibre bundle formulation may help attract the attention 
of researchers with different backgrounds.

\acknowledgments
This work was developed while D.V.\ was on a sabbatical leave
at the Institute for Quantum Optics and Quantum Information
(IQOQI-Vienna), of the Austrian Academy of Sciences. D.V.\ 
acknowledges
partial financial support from
the S\~ao Paulo State Research Foundation
(FAPESP) under grant no.\ 2023/04827-9. 
D.V.\ thanks IQOQI, the University 
of Vienna, and
\v Caslav Brukner and his group for the
stimulating environment and their extended hospitality. D.V.\ also thanks
Henrique Gomes 
for an 
interesting exchange on the subject, and Viktoria Kabel and 
Anne-Catherine de la Hamette for their kind attention and enlightening discussions about
Ref.~\cite{KHACGBB24}. J.B.\ also thanks IQOQI, \v{C}aslav Brukner and  Viktoria Kabel and 
Anne-Catherine de la Hamette, for their hospitality, kind attention and enlightening discussions. 

For corrections and comments on a previous version of this paper, we are very grateful to: Anne-Catherine de la Hamette,  Viktoria Kabel, Markus M\"uller, James Read, Eduardo Am\^ancio Oliveira, and Philipp H\"ohn; and especially to Henrique Gomes and
Jan G\l owacki, whose understanding of all the issues and whose helpfulness have been extraordinary.

\end{document}